\documentstyle[12pt]{article}
\def\fun#1#2{\lower3.6pt\vbox{\baselineskip0pt\lineskip.9pt
  \ialign{$\mathsurround=0pt#1\hfil##\hfil$\crcr#2\crcr\sim\crcr}}}
\newcommand{\dd}{\mbox{d}}
\newcommand{\vecc}[1]{\mbox{\boldmath $#1$}}
\newcommand{\Li}{\mbox{Li}_2}

\newcommand{\matr}[1]{\mbox{#1}}

\title { Violation of the factorization theorem in large--angle radiative
         Bhabha scattering }

\author {A.B.~Arbuzov, E.A.~Kuraev, and B.G.~Shaikhatdenov }

\date{}

\begin{document}

\maketitle

\begin{abstract}
The lowest order QED radiative corrections to the radiative
large angle Bhabha scattering process
in the region where all kinematic invariants
are large compared to the electron mass are considered.
We show that the leading logarithmic corrections do not factorize
before the Born cross section, contrary to the picture
assumed in the renormalization group approach.
The leading and non leading contributions
for typical kinematics of the
hard process at the energy of the $\Phi$ factory are estimated.
\end{abstract}

\section{Introduction}

The large angle Bhabha scattering process (LABS) plays an important role in
$e^+ e^-$ colliding beam physics~\cite{Dol}. First, it is traditionally
used for calibration, because it has a large cross section and can be
recognized easily. Second, it might provide essential background
information in a study of quarkonia physics. The result obtained
below can also be used to construct Monte Carlo event generators
for Bhabha scattering processes.

In our previous papers we considered the following contributions
to the large angle Bhabha cross section: pair production
(virtual, soft~\cite{VS}, and hard~\cite{r2}) and two hard photons~\cite{2h}.
This paper is devoted to the calculation of radiative corrections (RC)
to a single hard--photon emission process.
We consider the kinematics essentially of type $2 \to 3$,
in which all possible scalar products of 4 momenta of external particles
are large compared to the electron mass squared.

Considering virtual corrections, we identify gauge invariant sets of
Feynman diagrams (FD). Loop corrections associated with emission and
absorption of virtual photons by the same fermionic line are called
as Glass--type (G) corrections.
The case in which a loop involves exchange of two virtual photons
between different fermionic lines is called Box--type (B) FD.
The third class includes the vertex function and vacuum polarization
contributions ($\Gamma \Pi$--type). We see explicitly that all terms
that contain the square of large logarithms $\ln(s/m^2)$,
as well as those that contain the infrared singularity parameter
(fictitious photon mass $\lambda$), cancel out in the total
sum, where the emission of an additional soft photon is also considered.

We note here that the part of the general result associated with
scattering--type diagrams (see Fig.~1~(1,5)) was used to describe
radiative deep inelastic scattering (DIS) with RC taken into
account in Ref.~\cite{r1} (we labeled it the Compton tensor
with heavy photon). A Similar set of FD can be used to describe
the annihilation channel~\cite{r2}.

The problem of virtual RC calculations at the 1 loop level is
cumbersome for the process
\begin{equation}
e^+(p_2)\ +\ e^-(p_1)\ \longrightarrow\ e^+(p^{'}_2)\ +\ e^-(p^{'}_1)\
+\ \gamma(k_1).
\end{equation}
Specifically, if at the Born level we need to consider
eight FD, then at the 1 loop level we have as many as 72. Furthermore,
performing loop momentum integration, we introduce scalar, vector,
and tensor integrals up to the third rank with 2,3,4, and 5 denominators
(a set of relevant integrals is given in Appendices~A,~B).
A high degree of topological symmetry of FD for a cross section
can be exploited to calculate the matrix element squared.
Using them, we can restrict ourselves to the consideration
of interferences of the Born--level amplitudes (Fig.~1~(1-4))
with those that contain 1 loop integrals (Fig.~1~(5-16)).
Our calculation is simplified since
we omit the electron mass $m$ in evaluating the corresponding
traces due to the kinematic region under consideration:
\begin{eqnarray} \label{large}
&&s\sim s_1\sim -t_1\sim -t\sim - u\sim - u_1\sim \chi_{1,2}\sim
\chi^{'}_{1,2} \gg m^2,\\ \nonumber
&& s=2p_1p_2 \ , \quad t=-2p_2p'_2 \ , \quad u=-2p_1p'_2 \,  \quad
s_1=2p'_1p'_2 \ ,  \nonumber \\
&& t_1=-2p_1p'_1 \ , \quad u_1=-2p_2p'_1 \ ,\quad \chi_{1,2}=2k_1p_{1,2} \ ,
\quad \chi'_{1,2}=2k_1p'_{1,2} \ , \nonumber \\
&& s+s_1+t+t_1+u+u_1 = 0,\quad s+t+u=\chi_1', \nonumber \\ \nonumber
&&s _1+t+u_1=-\chi_1, \quad
 t + \chi_1=t_1+\chi_1'.
\end{eqnarray}

We found that some kind of local factorization took place both
for the G  and B type FD:
the leading logarithmic contribution to the
matrix element squared, summed over spin states,
arising from interference of one of the
four FD at the Born level (Fig.~1~(1-4)) with some 1 loop--corrected FD
(Fig.~1~(5-16)), turns out to be proportional to the interference of the
corresponding amplitudes at the Born level. The latter has the form
\begin{eqnarray} \label{Br}
E_0 &=& (4\pi\alpha)^{-3}\sum |M_1|^2=
 - \frac{16}{t^2}\;\frac{1}{4}\matr{Tr}(\hat{p}_1'O_{11'}
\hat{p}_1\tilde{O}_{11'})\cdot\;\frac{1}{4}\matr{Tr}(\hat{p}_2
\gamma_{\sigma}\hat{p}_2'\gamma_{\rho})   \nonumber \\
&=& - \frac{16}{t\chi_1\chi_1'}(u^2+u_1^2+s^2+s_1^2),\\
O_0 &=& (4\pi\alpha)^{-3}\sum M_1M_2^*
= \frac{8}{tt_1}\biggl( \frac{s}{\chi_1\chi_2}
+ \frac{s_1}{\chi_1'\chi_2'} + \frac{u}{\chi_1\chi_2'}
+ \frac{u_1}{\chi_2\chi_1'} \biggr)  \nonumber \\
&\times& (u^2+u_1^2+s^2+s_1^2), \nonumber \\
I_0 &=& (4\pi\alpha)^{-3}\sum M_1(M_3^*+M_4^*)=
-(1+\hat{Z})\frac{4}{ts_1}\biggl\{ - \frac{4u_1\chi_2'}{\chi_1} \nonumber\\
&+& \frac{4u(s_1+t_1)(s+t)}{\chi_2\chi_1'}
- \frac{2}{\chi_1\chi_2}[ 2suu_1 + (u+u_1)(uu_1+ss_1-tt_1)] \nonumber \\
&+& \frac{2}{\chi_1\chi_1'}[2t_1uu_1+(u+u_1)(uu_1
+ tt_1-ss_1)] \biggr\},\nonumber \\
O_{11'}&=&\gamma_{\rho}\frac{\hat p_1'+\hat k_1}{\chi_1'}\gamma_{\mu}
-\gamma_{\mu}\frac{\hat p_1-\hat k_1}{\chi_1}\gamma_{\rho}, \quad
\tilde{O}_{11'}=O_{11'}(\rho \leftrightarrow \mu),  \nonumber
\end{eqnarray}
where the $\hat{Z}$-operator acts as follows:
\begin{eqnarray*}
\hat Z= \left| \matrix{p_1 \longleftrightarrow p_1' & s
\longleftrightarrow s_1 \cr
p_2 \longleftrightarrow p_2' & u \longleftrightarrow u_1 \cr
k_1 \rightarrow -k_1 & t,t_1 \rightarrow t,t_1 \cr} \right|.
\end{eqnarray*}

It can be shown that the total matrix element
squared, summed over spin states, can be obtained using
symmetry properties realized by
means of the permutation operations:
\begin{eqnarray}
\sum |M|^2&=& (4\pi\alpha)^3F, \qquad
F=(1+\hat{P}+\hat{Q}+\hat{R})\Phi = \\
&=& 16\, \frac{ss_1(s^2+s_1^2)+tt_1(t^2+t_1^2) + uu_1(u^2+u_1^2)}
{ss_1tt_1} \nonumber \\
\qquad &\times& \biggl( \frac{s}{\chi_1\chi_2} + \frac{s_1}{\chi_1'\chi_2'}
- \frac{t}{\chi_2\chi_2'} - \frac{t_1}{\chi_1\chi_1'}
+ \frac{u}{\chi_1\chi_2'} + \frac{u_1}{\chi_2\chi_1'} \biggr), \nonumber \\
\Phi &=& E_0 + O_0 - I_0. \nonumber
\end{eqnarray}
The explicit form of the $\hat{P},\hat{Q},\hat{R}$ operators is
\begin{eqnarray}
\hat P &=& \left| \matrix{p_1 \longleftrightarrow -p_2' & s
\longleftrightarrow s_1 \cr
p_2 \longleftrightarrow -p_1' & t \longleftrightarrow t_1 \cr
k_1 \rightarrow k_1 & u,u_1 \rightarrow u,u_1 \cr} \right|, \nonumber \\
\hat Q &=& \left| \matrix{p_2 \longleftrightarrow -p_1' & s
\longleftrightarrow t_1 \cr
p_2' \rightarrow p_2' & s_1 \longleftrightarrow t \cr
p_1,k_1 \rightarrow p_1,k_1 & u,u_1 \rightarrow u,u_1 \cr} \right|, \\ \nonumber
\hat R &=& \left| \matrix{p_1 \longleftrightarrow -p_2' & s
\longleftrightarrow t \cr
p_1' \rightarrow p_1' & s_1 \longleftrightarrow t_1 \cr
p_2,k_1 \rightarrow p_2,k_1 & u,u_1 \rightarrow u,u_1 \cr} \right|.
\end{eqnarray}
The differential cross section at the Born level in the case of large angle
kinematics~(\ref{large}) was found in Ref.~\cite{ber}:
\begin{eqnarray}\label{Born}
\dd\sigma_0(p_1,p_2) &=& \frac{\alpha^3}{32s\pi^2}F
\frac{\dd^3p'_1\dd^3p'_2\dd^3k_1}{\varepsilon'_1\varepsilon'_2\omega_1}
\delta^{(4)}(p_1+p_2-p'_1-p'_2-k_1)\,,
\end{eqnarray}
where $\varepsilon_1$, $\varepsilon_2$, and $\omega_1$ are the energies
of the outgoing fermions and photon, respectively.
The  collinear kinematic regions (real photon emitted in
the direction of one of the charged particles) corresponding to the
case in which one of the invariants $\chi_i,\chi_i'$ is of order $m^2$
yields the main contribution to the total cross section. These
require separate investigation, and will be considered elsewhere.

Our paper is organized as follows. In Sec.~2 we consider
the contribution due to the set of FD Fig.~1~(5--8)
called {\em glasses\/} here (G type diagrams). Using
crossing symmetry, we construct the whole G type contribution from the
gauge--invariant set of FD in Fig.~1~(5). Moreover, only the
set of FD depicted
in Fig.~2~(d) can be considered in practical calculations,
due to an additional
mirror symmetry in the diagrams of Fig.~2~(d,~e). We therefore
start by checking the gauge invariance of the Compton tensor
described by the FD of Fig.~2~(d,~e)
for all fermions and one of the photons on the mass shell.
In Sec.~3 we consider the contribution of amplitudes containing vertex
functions and the virtual photon polarization operator shown in
Fig.~1~(13--15) and Fig.~2~(f,~d). In Sec.~4 we take into account the
contribution of FD with virtual two--photon exchange,
shown in Fig.~1~(9--12), called {\em boxes\/} here (B type diagrams).
Again, using the crossing symmetry
of FD, we show how to use only the FD of Fig.~1~(9) in calculations.
We show that the terms containing infrared singularities, as well as
thoes containing large logarithms, can be written in simple
form, related to certain contributions to the radiative Bhabha cross section
in the Born approximation~(\ref{Br}).
We also control terms in the matrix element squared that do not contain
large logarithms and are infrared--finite.
Thus our considerations permit us to calculate the cross section in the
kinematic region~(\ref{large}), in principle, to power--law accuracy,
i.e., neglecting terms that are
\begin{equation}\label{Acc}
{\cal O}\Biggl(\frac{\alpha}{\pi}\frac{m^2}{s}L_s^2\Biggr),
\end{equation}
as compared to ${\cal O}(1)$ terms calculated in this paper. Note
that the terms in~(\ref{Acc}) are less than $10^{-4}$ for typical
moderately high energy colliders (DA$\Phi$NE, VEPP-2M, BEPS).
Unfortunately, the non leading terms are too complicated
to be presented analytically, so we have estimation them numerically.
In Sec.~5 we consider emission of an additional soft photon in our
radiative Bhabha process.
To conclude, we note that the expression for the total correction,
taking into account virtual and real soft photon emission in the leading
logarithmic approximation, has a very elegant and handy form,
although it differs from what one might expect in the approach based on
renormalization group ideas.
Besides analytic expressions, we also give numerical values,
along with the non leading terms for a few points under
typical experimental conditions.

\section{Contribution of G type diagrams}

We begin by explicitly checking the gauge invariance of the  tensor
\begin{equation}
\bar u(p^{'}_1)R^{\sigma\mu}_{1,1'} u(p_1).
\end{equation}
This was done indirectly in Ref.~\cite{r1}, where the Compton
tensor for a
heavy photon was written in terms of explicitly gauge invariant tensor
structures. We use the expression
\begin{equation}
R^{\sigma\mu}_{1,1'} = R^{\chi_1} + R^{\chi^{'}_1}\, ,
\end{equation}
\begin{eqnarray}\label{rxi1}
R^{\chi_1} &=& A_2\gamma_{\sigma}\hat k_1\gamma_{\mu}
+ \int\frac{\dd^4k}{\matr{i}\pi^2} \Biggl\{ \frac
{\gamma_{\lambda}(\hat p^{'}_1-\hat k)\gamma_{\sigma}(\hat p_1-\hat k_1-
\hat k)\gamma_{\lambda}(\hat p_1-\hat k_1)\gamma_{\mu}}{-\chi_1(0)(2)(q)}
\nonumber \\
&+& \frac{\gamma_{\lambda}(\hat p^{'}_1-\hat k)\gamma_{\sigma}
(\hat p_1-\hat k_1-\hat k)\gamma_{\mu}(\hat p_1-\hat k)
\gamma_{\lambda}}{(0)(1)(2)(q)}\Biggr\},
\end{eqnarray}
where
\begin{eqnarray}\label{4}
(0)=k^2-\lambda^2, \quad
(2)=(p^{'}_1-k)^2-m^2, \quad (1)=(p_1-k)^2-m^2, \nonumber \\
(q)=(p_1-k_1-k)^2-m^2, \quad A_2=\frac{2}{\chi_1}\biggl(L_{\chi_1}-\frac{1}
{2}\biggr), \quad L_{\chi_1}=\ln\frac{\chi_1}{m^2}\, .
\end{eqnarray}

The quantity $R^{\chi_1}$ corresponds to the FD depicted in Fig.~2~(d),
while $R^{\chi_1'}$ corresponds to the FD in Fig.~2~(e).
The first term on the
right--hand side of Eq.~(\ref{rxi1}) corresponds to the first
two FD of Fig.~2~(d) under conditions~(\ref{large}).
The gauge invariance condition
$R^{\sigma\mu}_{1,1'}k_{\mu} =0$ is clearly satisfied.
The gauge invariance condition regarding
the heavy photon Lorentz index provides some check of the loop momentum
integrals, which can be found in Appendix~A:
\begin{equation}\label{5}
\bar u(p_1')R^{\sigma\mu}_{1,1'} u(p_1)q_{\sigma} e_{\mu}(k_1)=
A k_1^{\mu}e_{\mu}(k_1), \qquad
A = - 2\frac{L_{\chi_1}-2}{\chi_1} - 6\frac{L_{\chi_1'}-1}{\chi_1'}\, .
\end{equation}
The gauge invariance thus satisfied due to the Lorentz condition for
the on shell photon, $e(k_1)k_1=0$.
As stated above, the use of crossing symmetries of amplitudes permits us to
consider only $R^{\chi_1}$.
For interference of amplitudes at the Born level (see Fig.~1~(1--4)
and Fig.~1~(5--8)), we obtain in terms of the replacement operators
\begin{eqnarray} \label{6}
(\Delta|M|^2)_G=2^5\alpha^4\pi^2(1+\hat P+\hat Q+\hat R)(1+\hat
Z)[E_{15}^{\chi_1}+O_{25}^{\chi_1}-I_{35}^{\chi_1}-I_{45}^{\chi_1}],
\end{eqnarray}
with
\begin{eqnarray}
E_{15}^{\chi_1}&=&\frac{16}{t^2}\;\frac{1}{4}\matr{Tr}
(\hat p_1' R^{\chi_1}\hat p_1 O_{11'})\cdot \frac{1}{4}\matr{Tr}
(\hat p_2\gamma_{\rho}\hat p_2'\gamma_{\sigma}), \nonumber \\
O_{25}^{\chi_1} &=& \frac{16}{tt_1}\;\frac{1}{4}\matr{Tr}
(\hat p_1' R^{\chi_1}\hat p_1 \gamma_{\rho})\cdot \frac{1}{4}\matr{Tr}
(\hat p_2\gamma_{\sigma}\hat p_2'O_{22'}), \\
I_{35}^{\chi_1} &=& \frac{4}{ts_1}\;\frac{1}{4}\matr{Tr}
(\hat p_1' R^{\chi_1}\hat p_1 O_{12}\hat p_2\gamma_{\sigma}
\hat p_2'\gamma_{\rho}), \nonumber \\
I_{45}^{\chi_1} &=& \frac{4}{ts}\;\frac{1}{4}\matr{Tr}
(\hat p_1' R^{\chi_1}\hat p_1
\gamma_{\rho}\hat p_2\gamma_{\sigma}\hat p_2'O_{1'2'}), \nonumber \\
O_{11'}&=&\gamma_{\rho}\frac{\hat p_1'+\hat k_1}{\chi_1'}\gamma_{\mu}
-\gamma_{\mu}\frac{\hat p_1-\hat k_1}{\chi_1}\gamma_{\rho}, \nonumber \\
O_{22'} &=& \gamma_{\mu}\frac{-\hat p_2'-\hat k_1}{\chi_2'}\gamma_{\rho}
-\gamma_{\rho}\frac{-\hat p_2+\hat k_1}{\chi_2}\gamma_{\mu}, \nonumber \\
O_{12}&=&-\gamma_{\mu}\frac{\hat p_1-\hat k_1}{\chi_1}\gamma_{\rho}
- \gamma_{\rho}\frac{-\hat p_2+\hat k_1}{\chi_2}\gamma_{\mu}, \nonumber \\
O_{1'2'} &=& \gamma_{\rho}\frac{\hat p_1'+\hat k_1}{\chi_1'}\gamma_{\mu}
+\gamma_{\mu}\frac{-\hat p_2'-\hat k_1}{\chi_2'}\gamma_{\rho}. \nonumber
\end{eqnarray}
In the logarithmic approximation, the G type amplitude contribution to the
cross section has the form
\begin{eqnarray}
\dd \sigma_G &=& \frac{\dd \sigma_0}{F}\,\frac{\alpha}{\pi}
(1+\hat{P}+\hat{Q}+\hat{R})\Phi\biggl[-\frac{1}{2}L_{t_1}^2
+\frac{3}{2}L_{t_1}+2L_{t_1}\ln\frac{\lambda}{m}\biggr], \\ \nonumber
L_{t_1} &=& \ln\frac{-t_1}{m^2}.
\end{eqnarray}

\section{ Vacuum polarization and vertex insertion contributions}

Let us examine a set of $\Gamma \Pi$--type FD.
The contribution of the Dirac form factor of fermions and vacuum polarization
(see Fig.~3) can be parametrized as $(1+\Gamma_t)/(1-\Pi_t)$,
while the contribution of the Pauli form factor is proportional to the
fermion mass, and is omitted here. We obtain
\begin{equation}
\dd \sigma_{\Gamma\Pi}=\frac{\dd \sigma_0}{F}\,\frac{\alpha}{\pi}
2(1+\hat P+\hat Q+\hat R)(\Gamma_t+\Pi_t)\Phi,
\end{equation}
where
\begin{eqnarray}
&&\Gamma_t =\frac{\alpha}{\pi}\left\{\left(\ln\frac{m}{\lambda}-1\right)
\left(1-L_t\right)-\frac{1}{4}L_t-\frac{1}{4}L_t^2+\frac{1}{2}\zeta_2
\right\}, \\ \nonumber
&&\Pi_t = \frac{\alpha}{\pi}\left(\frac{1}{3}L_t-\frac{5}{9}\right)
\,, \qquad L_t=\ln\frac{-t}{m^2}.
\end{eqnarray}
In realistic calculations, the vacuum polarization due to hadrons and muons
can be taken into account in a very simple fashion~\cite{r4},
just by adding it to $\Pi_t$.

\section{Contribution of the B type set of Feynman diagrams}

A procedure resembling the one used in the previous section,
applied to the B type set of FD (Fig.~1~(9--12a)), enables us to use
only certain 1--loop diagrams in practical calculations,
specifically three of those
in the scattering channel with uncrossed exchanged photon legs:
\begin{eqnarray}\label{8}
(\Delta|M|^2)_B &=& 2^5\alpha^4\pi^2\; \matr{Re}\,(1+\hat P+\hat Q+\hat R)
\nonumber \\ &\times&
[(1-\hat P_{22'})I_{19}^{\chi_1}+(1+\hat P_{22'})I_{29}^{\chi_1} - I ],
\end{eqnarray}
where
\begin{eqnarray}
\hat P_{22'} &=& \left| \matrix{p_2 \longleftrightarrow -p_2' & s
\longleftrightarrow u  \cr
p_1 \longleftrightarrow p_1 & s_1 \longleftrightarrow u_1 \cr
p_1',k_1 \rightarrow p_1',k_1 & t,t_1 \rightarrow t,t_1 \cr} \right|\,,
\end{eqnarray}
and
\begin{eqnarray}
I_{19}^{\chi_1}&=&\int\frac{\dd^4k}{\matr{i}\pi^2}\frac{1}
{(0)(q)((p_2+k)^2-m^2)}
\frac{16}{t}\;\frac{1}{4}\matr{Tr}(\hat p_1' B^{\chi_1}\hat p_1 O_{11'})
\nonumber \\
&\times&\frac{1}{4}\matr{Tr}\,(\hat p_2\gamma_{\sigma}(-\hat p_2-\hat k)
\gamma_{\lambda}\hat p_2'\gamma_{\rho}),\nonumber \\
I_{29}^{\chi_1}&=&\int\frac{\dd^4k}{\matr{i}\pi^2}\frac{1}
{(0)(q)((p_2+k)^2-m^2)}
\frac{16}{t_1}\;\frac{1}{4}\matr{Tr}(\hat p_1' B^{\chi_1}\hat p_1
\gamma_{\rho}) \nonumber \\
&\times&\frac{1}{4}
\matr{Tr}\,(\hat p_2\gamma_{\sigma}(-\hat p_2-\hat k)
\gamma_{\lambda}\hat p_2'O_{22'}), \nonumber \\
I&=&\int\frac{\dd^4k}{\matr{i}\pi^2}\frac{1}{(0)(q)}\biggl\{
\frac{4}{s_1}\frac{1}{4}\matr{Tr}(\hat p_2'\gamma_{\rho}\hat p_1'
B^{\chi_1}\hat p_1O^{12}\hat p_2(\hat A+\hat B) \\ \nonumber
&+&\frac{4}{s_1}\frac{1}{4}\matr{Tr}(\hat p_2'O_{1'2'}\hat p_1
B^{\chi_1}\hat p_1\gamma_{\rho}\hat p_2(\hat A+\hat B) ) \biggr\}\,,\\
&&\hat A=\frac{\gamma_{\sigma}(-\hat p_2-\hat k)\gamma_{\lambda}}
{(p_2+k)^2-m^2}\,, \quad
\hat B=\frac{\gamma_{\lambda}(-\hat p_2'+\hat k)\gamma_{\sigma}}
{(-p_2'+k)^2-m^2}. \nonumber
\end{eqnarray}
Here
\begin{eqnarray}
B^{\chi_1} &=&\frac{\gamma_{\lambda}(\hat p_1-\hat k_1-\hat k)
\gamma_{\sigma}(\hat p_1-\hat k_1)\gamma_{\mu}}{-\chi_1(d)}
+ \frac{\gamma_{\lambda}(\hat p_1-\hat k_1-\hat k)\gamma_{\mu}(\hat p_1-
\hat k)\gamma_{\sigma}}{(d)(1)} \nonumber \\
&+&\frac{\gamma_{\mu}(\hat p_1'+\hat k_1)\gamma_{\lambda}(\hat p_1-\hat k)
\gamma_{\sigma}}{\chi_1'(1)} \,,\quad (q)=(p_2-p_2'+k)^2-\lambda^2, \\
(d)&=&(p_1-k_1-k)^2-m^2, \ \ (1)=(p_1-k)^2-m^2, \ \ (0)=k^2-\lambda^2.
\nonumber
\end{eqnarray}
Analytic evaluations divulge a lack
of both double logarithmic ($\sim L_s^2$) and infrared
logarithmic ($\sim \ln(\lambda/m)L$) terms in the box
contribution. In spite of the explicit proportionality of the individual
contributions to the structures $E_0$, $O_0$, and $I_0$, the overall expression
terns out to be somewhat convoluted, despite its having a
factorized form in each
gauge--invariant subset of diagrams.
We parametrize the correction coming from the B type FD as follows:
\begin{eqnarray}
\dd \sigma_B=\dd \sigma_0\frac{\alpha}{\pi}L_s \Delta_B, \qquad
\Delta_B=2\ln\frac{ss_1}{uu_1}+\frac{2}{F}(\Phi_Q+\Phi_R)
\ln\frac{tt_1}{ss_1}\, .
\end{eqnarray}
The total virtual correction to the cross section has the form
\begin{eqnarray} \label{virt}
\dd \sigma^{\mathrm{virt}} &=& \dd \sigma_G + \dd \sigma_{\Gamma\Pi}
+ \dd \sigma_B = \frac{\alpha}{\pi} \biggl[ -L_s^2
\nonumber \\
&+& L_s\biggl(\frac{11}{3}+4\ln\frac{\lambda}{m}
+\Delta_G+\Delta_{\Gamma\Pi}+\Delta_B\biggr)
+{\cal{O}}(1)\biggr], \\ \nonumber
\Delta_G+\Delta_{\Gamma\Pi}&=&\frac{1}{F}\biggl(\Phi\ln\frac{s^2}{tt_1}
+ \Phi_R\ln\frac{t^2}{ss_1}+\Phi_Q\ln\frac{t_1^2}{ss_1}
+\Phi_P\ln\frac{s_1^2}{tt_1}\biggr),
\end{eqnarray}
where $\Phi_P=\hat P \Phi$, $\Phi_Q=\hat Q \Phi$, and $\Phi_R=\hat R \Phi$.

\section{Contribution from additional soft photon emission}

Consider now radiative Bhabha scattering accompanied by
emission of an additional soft photon in the center of mass
reference frame. By {\it soft} we mean that its energy does
not exceed some small quantity $\Delta\varepsilon$,
compared to the energy $\varepsilon$ of the initial beams.
The corresponding cross section has the form
\begin{eqnarray} \label{dsoft}
\dd\sigma^{\mathrm{soft}} &=& \dd\sigma_0\cdot\delta^{\mathrm{soft}},
\\ \nonumber
\delta^{\mathrm{soft}} &=& - \frac{4\pi\alpha}{16\pi^3}
\int\frac{\dd^3k_2}{\omega_2}\Biggl(
-\frac{p_1}{p_1k_2} + \frac{p^{'}_1}{p^{'}_1k_2} + \frac{p_2}{p_2k_2}
- \frac{p^{'}_2}{p^{'}_2k_2}\Biggr)^2\Bigg|_{\omega_2<\Delta\varepsilon}.
\end{eqnarray}

The soft photon energy does not exceed $\Delta\varepsilon \ll \varepsilon_1 =
\varepsilon_2 \equiv \varepsilon \sim \varepsilon^{'}_1
\sim \varepsilon^{'}_2$. In order
to calculate the right--hand side of Eq.~(\ref{dsoft}), we use the
master equation~\cite{r5}:
\begin{eqnarray}
&& - \frac{4\pi\alpha}{16\pi^3}\int\frac{\dd^3k}{\omega}
\frac{(q_i)^2}{(q_ik)^2}\Bigg|_{\omega<\Delta\varepsilon} =
- \frac{\alpha}{\pi}\ln\Biggl(\frac{\Delta
\varepsilon\cdot m}{\lambda\cdot\varepsilon_i}\Biggr)\ , \quad \omega = \sqrt
{k^2+\lambda^2}\, , \\ \label{4.3}
&&\frac{4\pi\alpha}{16\pi^3}\int\frac{\dd^3k}{\omega}\frac{2q_1q_2}{(kq_1)
(kq_2)}\Bigg|_{\omega<\Delta\varepsilon} = \frac{\alpha}{\pi}\Biggl[L_q\ln\Biggl(
\frac{m^2(\Delta\varepsilon)^2}{\lambda^2\varepsilon_1\varepsilon_2}\Biggr)
+ \frac{1}{2}L^2_q  \nonumber\\
&&\qquad - \frac{1}{2}\ln^2\biggl(\frac{\varepsilon_1}{\varepsilon_2}\biggr)
- \frac{\pi^2}{3} + \Li\left(\cos^2\frac{\theta}{2}\right)
\Biggr].
\end{eqnarray}
Here we used the notation
\begin{eqnarray}
&&L_q = \ln\frac{-q^2}{m^2}\, , \quad q_1^2 = q_2^2 = m^2, \quad
-q^2 = -(q_1-q_2)^2 \gg m^2, \\
&& q_{1,2} = (\varepsilon_{1,2},\vecc q_{1,2}),
\qquad \theta = \widehat{\vecc q_1\vecc q_2}, \nonumber
\end{eqnarray}
where $\varepsilon_1$, $\varepsilon_2$, and $\theta$ are the energies and
angle
between the 3 momenta $\vecc q_1, \vecc q_2,$ respectively,
and $\lambda $ is the
fictitious photon mass (all defined in the center of mass system).

The contributions of each possible term
on the right--hand side of Eq.~(\ref{dsoft}) are
\begin{eqnarray}
\frac{\pi}{\alpha}\delta^{\mathrm{soft}} &=& - \Delta_1 - \Delta_2
- \Delta_1' - \Delta_2' + \Delta_{12} + \Delta_{1'2'} + \Delta_{11'}
+ \Delta_{22'} \nonumber \\
&-& \Delta_{12'} - \Delta_{1'2}\, , \nonumber \\
\Delta_1 \!\! &=& \!\! \Delta_2 = \ln\frac{\Delta\varepsilon\cdot m}{\varepsilon\lambda}
\, , \quad \Delta_1' = \ln\frac{\Delta\varepsilon\cdot m}{\varepsilon^{'}_1
\lambda}\, , \quad \Delta_2' = \ln\frac{\Delta\varepsilon\cdot m}{\varepsilon^{'}_2\lambda}\, , \nonumber \\
\Delta_{12} \!\! &=& \!\! 2L_s\ln\frac{\Delta\varepsilon\cdot m}
{\varepsilon\lambda} + \frac{1}{2}L_s^2 - \frac{\pi^2}{3}\, ,\\
\Delta_{1'2'} \!\! &=& \!\! L_{s_1}\ln\Biggl(\frac{(\Delta\varepsilon\cdot m)^2}
{\varepsilon^{'}_1\varepsilon^{'}_2\lambda^2}\Biggr) + \frac{1}{2}L_{s_1}^2
-\frac{1}{2}\ln^2\Biggl(\frac{\varepsilon^{'}_1}{\varepsilon^{'}_2}\Biggr)
- \frac{\pi^2}{3} + \Li\left(\cos^2\frac{\theta_{1'2'}}{2}\right), \nonumber \\
\Delta_{11'} \!\! &=& \!\! L_{t_1}\ln\Biggl(\frac{(\Delta\varepsilon\cdot m)^2}
{\varepsilon^{'}_1\varepsilon\lambda^2}\Biggr) + \frac{1}{2}L_{t_1}^2
-\frac{1}{2}\ln^2\Biggl(\frac{\varepsilon^{'}_1}{\varepsilon}\Biggr)
- \frac{\pi^2}{3} + \Li\left(\cos^2\frac{\theta_{1'}}{2}\right), \nonumber \\
\Delta_{22'} \!\! &=& \!\! L_{t}\ln\Biggl(\frac{(\Delta\varepsilon\cdot m)^2}
{\varepsilon\varepsilon^{'}_2\lambda^2}\Biggr) + \frac{1}{2}L_{t}^2
-\frac{1}{2}\ln^2\Biggl(\frac{\varepsilon^{'}_2}{\varepsilon}\Biggr)
- \frac{\pi^2}{3} + \Li\left(\sin^2\frac{\theta_{2'}}{2}\right), \nonumber \\
\Delta_{1'2} \!\! &=& \!\! L_{u_1}\ln\Biggl(\frac{(\Delta\varepsilon\cdot m)^2}
{\varepsilon\varepsilon^{'}_1\lambda^2}\Biggr) + \frac{1}{2}L_{u_1}^2
-\frac{1}{2}\ln^2\Biggl(\frac{\varepsilon^{'}_1}{\varepsilon}\Biggr)
- \frac{\pi^2}{3} + \Li\left(\sin^2\frac{\theta_{1'}}{2}\right), \nonumber \\
\Delta_{12'} \!\! &=& \!\! L_{u}\ln\Biggl(\frac{(\Delta\varepsilon\cdot m)^2}
{\varepsilon\varepsilon^{'}_2\lambda^2}\Biggr) + \frac{1}{2}L_{u}^2
-\frac{1}{2}\ln^2\Biggl(\frac{\varepsilon}{\varepsilon^{'}_2}\Biggr)
- \frac{\pi^2}{3} + \Li\left(\cos^2\frac{\theta_{2'}}{2}\right), \nonumber \\
&&L_u=\ln\frac{-u}{m^2}\,,\quad L_{u_1}=\ln\frac{-u_1}{m^2}\,,\quad
\Li(z) \equiv - \int\limits_{0}^{z}\frac{\dd x}{x}\ln(1-x),  \nonumber
\end{eqnarray}
where $\varepsilon^{'}_1, \varepsilon^{'}_2$ are the center of mass
energies of the scattered electron and positron, respectively; $\theta_{1'}, \theta_{2'}$
are their scattering angles (measured from the initial electron momentum
direction); and $\theta_{1'2'}$ is the angle between the scattered electron and
positron momenta.

Separating out large logarithms, we obtain
\begin{eqnarray} \label{soft}
&& \!\!\!\! \delta^{\mathrm{soft}} = \frac{\alpha}{\pi} \biggl\{
4(L_s - 1)\ln\frac{m\Delta\varepsilon}{\lambda\varepsilon}
+ L_s^2 + L_s\ln\frac{tt_1}{uu_1} + L_s\ln\frac{1-c_{1'2'}}{2}
+{\cal O}(1) \biggr\},  \nonumber \\
&& \qquad c_{1'2'}= \cos\theta_{1'2'}.
\end{eqnarray}
This can be written in another form, using experimentally measurable
quantities, the relative energies of the scattered leptons and the scattering
angles:
\begin{eqnarray} \label{eq:31}
y_i &=& \frac{\varepsilon_i'}{\varepsilon}\,,\quad c_i=\cos\theta_i'\,,
\quad \frac{1}{2}(1-c_{1'2'})=\frac{y_1+y_2-1}{y_1y_2}\,,
\nonumber \\
-\frac{t}{s}&=& y_2\frac{1+c_2}{2}\,, \
- \frac{u}{s}=y_2\frac{1-c_2}{2} \,,\
-\frac{t_1}{s}=y_1\frac{1-c_1}{2}\,, \\ \nonumber
\frac{s_1}{s} &=& y_1+y_2-1 \,,\
- \frac{u_1}{s}=y_1\frac{1+c_1}{2}.
\end{eqnarray}

\section{Conclusions}

The double logarithmic terms of type $L_s^2$ and
those proportional to $L_s\ln(\lambda/m)$ cancel in the overall sum
with the corresponding terms from the soft photon contribution ~(\ref{soft}).
Omitting vacuum polarization, we obtain in
the logarithmic approximation
\begin{eqnarray} \label{sv}
\dd \sigma^{\mathrm{soft+virt}}&=&\dd \sigma_0\frac{\alpha}{\pi}\biggl[
L_s\biggl(4\ln\frac{\Delta\varepsilon}{\varepsilon}+\Delta_L\biggr)
+\Delta(y_1,y_2,c_1,c_2)\biggr], \\
\Delta_L&=&3+\ln\frac{(1-c_1)(1-c_2)}{(1+c_1)
(1+c_2)}+\ln\frac{y_1+y_2-1}{y_1y_2} \nonumber \\
&+& \frac{1}{F}\biggl[
\Phi\ln\frac{s^2}{tt_1}+\Phi_P\ln\frac{s_1^2}{tt_1}
+ \Phi_Q\ln\frac{t_1^2}{ss_1}
+ \Phi_R\ln\frac{t^2}{ss_1}\biggr] \nonumber \\ \nonumber
&+& 2\ln\frac{ss_1}{uu_1}
+\frac{2}{F}(\Phi_Q+\Phi_R)\ln\frac{tt_1}{ss_1}.
\end{eqnarray}

The function $\Delta(y_1,y_2,c_1,c_2)$ is quite complicated.
To compare it with $\Delta_L$, we give their numerical values
(omitting vacuum polarization)
for a certain set of points from physical regions~(\ref{reg}) and
$y_1+y_2 > 1$, $D > 0$ (see Table~1).
Considering the kinematics typical of
large angle inelastic Bhabha scattering, we show the lowest--order
contribution
previously obtained~\cite{EKP} and the radiative corrections
calculated in this work.

\begin{table}
\begin{center}
\begin{tabular}{|c||c|c|c|c||r|r|}
\hline
${\cal N}$&$y_1$&$y_2$&$c_1$&$c_2$&$\Delta_L$&$\Delta$
\\ \hline
$1$&$0.36$&$0.89$&$-0.70$&$-0.10$&$10.70$&$-24.53$ \\ \hline
$2$&$0.59$&$0.66$&$0.29$&$-0.06$&$4.86$&$-11.41$ \\ \hline
$3$&$0.67$&$0.67$&$0.50$&$0.30$&$5.82$&$-35.58$ \\ \hline
$4$&$0.68$&$0.65$&$0.60$&$-0.50$&$4.10$&$-10.45$ \\ \hline
\end{tabular}
\end{center}
\label{num}
\caption{ Numerical estimates of $\Delta_L\ {\matr{and}}\ \Delta$
          versus $y_1, y_2, c_1, c_2$}
\end{table}

After performing loop integration and shifting logarithms
$(L_i = L_s + L_{is})$, one can see that the terms containing infrared
singularities and double logarithmic terms $\sim L_s^2$, are associated
with a factor equal to the corresponding Born contribution.
This is true of all types of contributions.

The phase volume
$$
\dd \Gamma =
\frac{\dd^3p'_1\dd^3p'_2\dd^3k_1}{\varepsilon'_1\varepsilon'_2\omega_1}
\delta^{(4)}(p_1+p_2-p'_1-p'_2-k_1)
$$
can be transformed in various ways~\cite{EKP}. We introduce the variables
(see Eq.~(\ref{eq:31}))
\begin{equation} \label{reg}
y_i=\frac{\varepsilon_i'}{\varepsilon}\,,\quad c_i=\cos\theta_i'\,,
\quad \theta_i'=\widehat{\vecc{p}_1,\vecc{p}}_i'\,,\quad 0<y_i<1\,,\quad
-1<c_{1,2}<1,
\end{equation}
which parametrize the kinematics of the outgoing particles (these do not
include a common degree of freedom, a rotation about the beam axis).
The phase volume then takes the form
\begin{eqnarray}
&&\dd\Gamma=\frac{\pi s\dd y_1\dd y_2\dd c_1\dd c_2 }{2\sqrt{D(y_1,y_2,c_1,c_2)}}
\Theta(y_1+y_2-1)\Theta(D(y_1,y_2,c_1,c_2)), \nonumber \\
&&D(y_1,y_2,c_1,c_2)=\rho^2 - c_1^2 - c_2^2 - 2c_{1'2'}c_1c_2,\\ \nonumber
&& \rho^2 = 2(1-c_{1'2'})\frac{(1-y_1)(1-y_2)}{y_1y_2}\, .
\end{eqnarray}
The allowed region of integration is a triangle in the $y_1,y_2$ plane
and the interior of the ellipse $D>0$ in the $c_1,c_2$ plane.

We now discuss the relation of our result to the renormalization
group approach.
The dependence on $\Delta\varepsilon/\varepsilon$ in~(\ref{sv})
disappears when one takes into account hard two--photon emission.
The leading contribution arises from the kinematics when the second
hard photon is emitted close to the direction of motion of one of
the incoming or outgoing particles:
\begin{eqnarray} \label{hard}
\dd \sigma^{\mathrm{hard}} &=& \frac{\alpha}{2\pi}L_s
\biggl[\frac{1+z^2}{1-z}\biggl(\dd \sigma_0(zp_1,p_2,p_1',p_2')
+\dd \sigma_0(p_1,zp_2,p_1',p_2')\biggr)\dd z \nonumber \\
&+& \frac{1+z_1^2}{1-z_1}
\dd \sigma_0\left(p_1,p_2,\frac{p_1'}{z_1},p_2'\right)\dd z_1
+\frac{1+z_2^2}{1-z_2}
\dd \sigma_0\left(p_1,p_2,p_1',\frac{p_2'}{z_2}\right)\dd z_2\biggr], \nonumber\\
&&z=1-x_2\,,\quad z_i=\frac{y_i}{y_i+x_2}\,,\quad x_2=\frac{\omega_2}
{\varepsilon}.
\end{eqnarray}
The fractional energy of the additional
photon varies within the limits $\Delta\varepsilon/\varepsilon<x_2
=\omega_2/\varepsilon<1$.
This formula agrees with the Drell--Yan form of radiative Bhabha
scattering (with switched--off vacuum polarization)
\begin{eqnarray} \label{DY}
\dd \sigma(p_1,p_2,p_1',p_2')\!\! &=& \!\! \int \!
\dd x_1\,\dd x_2\,{\cal{D}}(x_1){\cal{D}}(x_2)
\,\dd \sigma_0\,\!\!\left(x_1p_1,x_2p_2,\frac{p_1'}{z_1},
\frac{p_2'}{z_2}\right) \nonumber \\ &\times&
{\cal{D}}(z_1){\cal{D}}(z_2)\dd z_1\dd z_2,
\end{eqnarray}
where the non singlet structure functions ${\cal{D}}$ are~\cite{KF}
\begin{eqnarray}
{\cal{D}}(z)&=&\delta(1-z)+\frac{\alpha}{2\pi}L{\cal{P}}^{(1)}(z)
+\biggl(\frac{\alpha}{2\pi}L\biggr)^2\frac{1}{2!}\,
{\cal{P}}^{(2)}(z)+ \ldots\,,
\nonumber \\
{\cal{P}}^{(1)}(z)&=&\lim_{\Delta \to 0}\left[
\frac{1+z^2}{1-z}\Theta(1-z-\Delta)+\delta(1-z)
\left(2\ln\Delta+\frac{3}{2}\right)\right].
\end{eqnarray}

In our calculations we see explicitly a factorization of the terms
containing double logarithmic contributions and infrared single
logarithmic ones, which arise from G  and $\Gamma\Pi$ type FD.
To be precise, the
corresponding contributions to the cross section have the structure of the
Born cross section~(\ref{Born}). But the above claim
fails to be true for terms containing single logarithms.
Hence, the Drell--Yan form~(\ref{DY}) is not valid in this case,
and the factorization theorem breaks down, because the mass
singularities (large logarithms) do not factorize before the
Born structure. That is because of plenty of different type
amplitudes and kinematic variables, which describe our process.
The reason for the violation of a naive usage of factorization
in the Drell--Yan form has presumably the same origin with that
found in Ref.~\cite{SG}, where the authors claimed that it
is necessary to study independently the renormalization group
behavior of leading logarithms before different amplitudes of
the same process. Note that in the
$e\mu \to e\mu\gamma$ reaction, which can easily be extracted
from our results, factorization does take place.
We also see from~(\ref{sv}) that factorization will take place
if all the logarithmic terms become equal, i.e.,
$\ln(s_1/m^2)=\ln(s/m^2)=\ldots$\ .
The source for the violation of the factorization theorem, we found,
might have a relation to some of those found in other
problems~\cite{viol}.

Numerical estimates (see Table 1) for the $\Phi$ factory energy range
($\sqrt s\simeq 1\ {\matr{GeV}}$) shows that the contribution of
the non leading terms coming from virtual and soft real photon emission
might reach $35\%$.
Additional
hard photon emission will also contribute to $\Delta_L$ and $\Delta$.
To get an explicit form of that correction, one has to take
into account a definite experimental setup.

Obviously, an analogous phenomenon of the factorization theorem
violation takes place in QCD in processes like $q\bar{q}\to q\bar{q}g$
and $q\bar{q}\to q\bar{q}\gamma$.
A consistent investigation of the latter processes,
taking into account the phenomenon found,
can give a certain correction to predictions for large angle
jet production and direct hard photon emission at
proton--antiproton colliders.

\subsection*{Acknowledgments}

We are grateful to D.~V.~Shirkov for useful discussions and
pointing~\cite{SG} out to us.
We also thank A.~Belitsky, P.~Ferro, N.~Merenkov and
L.~Trentadue for participating at the very beginning of this investigation.
We are also indebted to A.~Belitsky for assistance in the creation
of the tables of integrals~\cite{Table} used in this paper.
This work was supported in part by INTAS grant 93--1867 ext.

\section*{Appendix A}
{\large \bf Loop integrals for G--type Feynman diagrams}
\vskip 20.0pt
\setcounter{equation}{0}
\renewcommand{\theequation}{A.\arabic{equation}}

In this and the following appendices we used partially the results
of our previous work~\cite{Table} and refer to it for further details.
After this digression let us turn to the problem.
Two types of FD require different approaches.
For the set of FD, labeled as {\em glasses\/} (G), only three
independent external momenta are relevant due to the conservation
law: $p_1 + q = p_1' + k_1$. Choosing $p_1$, $p_1'$, $q$ we use the notation:
\begin{eqnarray}
J_{ijk} &=& \int\frac{\dd^4 k}{\matr{i}\pi^2}\frac{1}{(i)(j)(k)}\, ,\qquad
J_{012q} = \int\frac{\dd^4 k}{\matr{i}\pi^2}\frac{1}{(0)(1)(2)(q)}\, ,
\nonumber \\
J_{ijk}^{\mu} &=& \int\frac{\dd^4 k}{\matr{i}\pi^2}\frac{k^{\mu}}{(i)(j)(k)}
= a_{ijk}p_1^{\mu} + b_{ijk}p_1'^{\mu} +c_{ijk}q^{\mu}, \nonumber \\
J_{ij\dots}^{\mu\nu} &=& \int\frac{\dd^4 k}{\matr{i}\pi^2}
\frac{k^{\mu}k^{\nu}}{ij\dots}
= g^T_{ij\dots}g^{\mu\nu} + a^T_{ij\dots}p_1^{\mu}p_1^{\nu}
+ b^T_{ij\dots}p_1'^{\mu}p_1'^{\nu} + c^T_{ij\dots}q^{\mu}q^{\nu} \\
&+& \alpha^T_{ij\dots}(p_1p_1')^{\mu\nu} + \beta^T_{ij\dots}(p_1q)^{\mu\nu}
+ \gamma^T_{ij\dots}(p_1'q)^{\mu\nu}, \nonumber \\
J_{012q}^{\mu\nu\lambda} &=& \int\frac{\dd^4 k}{\matr{i}\pi^2}
\frac{k^{\mu}k^{\nu}k^{\lambda}}{(0)(1)(2)(q)}
= K_{g1}(gp_1)^{\mu\nu\lambda} + K_{g2}(gp_1')^{\mu\nu\lambda}
+ K_{gq}(gq)^{\mu\nu\lambda} \nonumber \\
&+& K_{111}p_1^{\mu}p_1^{\nu}p_1^{\lambda}
+ K_{222}p_1'^{\mu}p_1'^{\nu}p_1'^{\lambda}
+ K_{qqq}q^{\mu}q^{\nu}q^{\lambda}
+ K_{112}(p_1^2p_1')^{\mu\nu\lambda} \nonumber \\ \label{defint}
&+& K_{122}(p_1p_1'^2)^{\mu\nu\lambda}
+ K_{11q}(p_1^2q)^{\mu\nu\lambda}
+ K_{1qq}(p_1q^2)^{\mu\nu\lambda}
+ K_{22q}(p_1'^2q)^{\mu\nu\lambda}  \nonumber \\
&+& K_{2qq}(p_1'q^2)^{\mu\nu\lambda}
+ K_{12q}(p_1p_1'q)^{\mu\nu\lambda}, \nonumber
\end{eqnarray}
where the inverse propagators are
\begin{eqnarray}
&&(0) = k^2 - \lambda^2, \ (1) = (p_1-k)^2-m^2, \nonumber \\
&&(2) = (p_1'-k)^2-m^2, \ (q) = (p_1'-q-k)^2 - m^2,
\end{eqnarray}
$\lambda$ is a fictitious photon mass. The symmetrized tensor
structures are defined as follows:
\begin{eqnarray*}
(pq)^{\mu\nu} &=& p^{\mu}q^{\nu} + p^{\nu}q^{\mu},\qquad
(p^2q)^{\mu\nu\lambda} = p^{\mu}p^{\nu}q^{\lambda} + p^{\mu}p^{\lambda}q^{\nu}
+ p^{\nu}p^{\lambda}q^{\mu}, \\
(gp)^{\mu\nu\rho} &=& g^{\mu\nu}p^{\rho}+g^{\mu\rho}p^{\nu}+g^{\rho\nu}p^{\mu},
\\
(pqr)^{\mu\nu\lambda} &=& p^{\mu}q^{\nu}r^{\lambda}
+ p^{\mu}q^{\lambda}r^{\nu} + p^{\nu}q^{\mu}r^{\lambda}
+ p^{\nu}q^{\lambda}r^{\mu} + p^{\lambda}q^{\mu}r^{\nu}
+ p^{\lambda}q^{\nu}r^{\mu}.
\end{eqnarray*}

The vector and tensor integrals can be calculated by multiplying
both sides of the expression~(\ref{defint}) by vectors $p_1^{\mu}$,
$p_1'^{\mu}$ and $q^{\mu}$. Then one has to use the relations
\begin{eqnarray}
2p_1k = (0) - (1), \quad 2k_1k = (q) - (1) + \chi_1, \quad
2p_1'k = (0) - (2),
\end{eqnarray}
and compare the coefficients before vector components on both sides.

Considering the vector and tensor integrals with three denominators,
we use ultra--violet divergent integrals with two denominators. Using
the Feynman trick to join denominators, they can be expressed as
\begin{eqnarray}
&& \int\frac{\dd^4 k}{\matr{i}\pi^2} \frac{1}{[(k-b)^2-d]^2}
= \ln\frac{\Lambda^2}{d} - 1, \nonumber \\
&& \int\frac{\dd^4 k}{\matr{i}\pi^2} \frac{k^{\mu}}{[(k-b)^2-d]^2}
= b^{\mu}\biggl( \ln\frac{\Lambda^2}{d} - \frac{3}{2}\biggr).
\end{eqnarray}
We put here the complete list of these integrals (in
approximation Eq.(\ref{large})):
\begin{eqnarray}
J_{01} &=& L_{\Lambda} + 1, \quad J_{1q} = L_{\Lambda} - 1, \quad
J_{2q} = L_{\Lambda} - L_t + 1, \nonumber \\
J_{0q} &=& L_{\Lambda} - L_{\chi_1} + 1, \quad
J_{12} = L_{\Lambda} - L_{t_1} + 1, \quad
J_{02} = L_{\Lambda} + 1, \nonumber \\
J_{01}^{\mu} &=& \frac{1}{2}p_1^{\mu}\biggl(L_{\Lambda}-\frac{1}{2}\biggr),
\quad
J_{1q}^{\mu} = (p_1^{\mu}-\frac{1}{2}k_1^{\mu})
\biggl(L_{\Lambda}-\frac{3}{2}\biggr), \nonumber \\
J_{2q}^{\mu} &=& \frac{1}{2}(p_1^{\mu}-k_1^{\mu}+p_1'^{\mu})
\biggl(L_{\Lambda}-L_t+\frac{1}{2}\biggr), \quad
J_{0q}^{\mu} = (p_1^{\mu}-k_1^{\mu})
\biggl(\frac{1}{2}L_{\Lambda}-\frac{1}{2}L_{\chi_1}+\frac{1}{4}\biggr),
\nonumber \\
J_{12}^{\mu} &=& (p_1^{\mu}+p_1'^{\mu})
\biggl(\frac{1}{2}L_{\Lambda}-\frac{1}{2}L_{t_1}+\frac{1}{4}\biggr), \quad
J_{02}^{\mu} = p_1'^{\mu}\biggl(\frac{1}{2}L_{\Lambda}-\frac{1}{4}\biggr),
\end{eqnarray}
where
\begin{eqnarray*}
L_q = L_t = \ln\frac{-t}{m^2},\quad L_{\chi_1} = \ln\frac{\chi_1}{m^2}, \quad
L_{\chi_1'} = \ln\frac{\chi_1'}{m^2} - {\matr{i}}\pi, \quad
L_{\Lambda} = \ln\frac{\Lambda^2}{m^2}.
\end{eqnarray*}

The scalar integrals with three denominators read
\begin{eqnarray}
J_{012} &=& \frac{1}{2t_1}\left[-2L_{\lambda}L_{t_1}+L_{t_1}^2
-\frac{\pi^2}{3}\right], \qquad
J_{12q} = \frac{1}{2(\chi_1'-\chi_1)}(L_{t}^2-L_{t_1}^2), \nonumber \\
J_{02q} &=& \frac{1}{t+\chi_1}\left[L_t(L_t-L_{\chi_1})
+ \frac{1}{2}(L_t-L_{\chi_1})^2 + 2\Li\left(1+\frac{\chi_1}{t}\right)\right],
\nonumber \\
J_{01q} &=& -\frac{1}{2\chi_1}L_{\chi_1}^2 - \frac{\pi^2}{3\chi_1}\, , \quad
\Li(z) \equiv - \int\limits_{0}^{z}\frac{\dd x}{x}\ln(1-x), \quad
L_{\lambda} = \ln\frac{\lambda^2}{m^2}\, .
\end{eqnarray}
The coefficients for vector integrals with three denominators are
\begin{eqnarray}
a_{012} &=& b_{012} = \frac{1}{t_1}L_{t_1}, \quad c_{012} = 0, \nonumber \\
a_{01q} &=& J_{01q} + \frac{2}{\chi_1}(L_{\chi_1}-1),\quad
b_{01q} = - c_{01q} = \frac{1}{\chi_1}(-L_{\chi_1}+2), \nonumber \\
a_{02q} &=& 0,\quad b_{02q} = \frac{\chi_1}{\chi_1+t}J_{02q} + \frac{2tL_t}{(\chi_1+t)^2}
+ \frac{(\chi_1-t)L_{\chi_1}}{(\chi_1+t)^2}\, ,\quad
c_{02q} = \frac{L_{\chi_1} - L_t}{\chi_1+t}\, , \nonumber \\
a_{12q} &=& \frac{t}{t-t_1}J_{12q} + \frac{(t+t_1)L_{t_1}-2tL_t}{(t-t_1)^2}
+ \frac{2}{t-t_1}\, ,\quad
b_{12q} = J_{12q}-a_{12q}\, ,\nonumber \\
c_{12q} &=& \frac{t_1}{t-t_1}J_{12q} + \frac{-(t+t_1)L_t+2t_1L_{t_1}}{(t-t_1)^2}
+ \frac{2}{t-t_1}.
\end{eqnarray}
The tensor integrals for G--type FD (see Eq.(A.1)) have the following form:
\begin{eqnarray}
g^T_{012} &=& \frac{1}{4}(L_{\Lambda}-L_{t_1})+\frac{3}{8}\, , \quad
a^T_{012}=b^T_{012}=\frac{1}{2t_1}(L_{t_1}-1), \quad
\alpha^T_{012}=\frac{1}{2t_1}\, , \nonumber \\
c^T_{012} &=& \beta^T_{012}=\gamma^T_{012}=0,
\end{eqnarray}

\begin{eqnarray*}
g^T_{01q} &=& \frac{1}{4}(L_{\Lambda}-L_{\chi_1})+\frac{3}{8}\, , \quad
a^T_{01q} = J_{01q}+\frac{3}{\chi_1}L_{\chi_1}-\frac{9}{2\chi_1}\, , \\
b^T_{01q} &=& c^T_{01q}=-\gamma^T_{01q}=-\frac{1}{2\chi_1}(L_{\chi_1}-2), \quad
\beta^T_{01q} = -\alpha^T_{01q} = \frac{1}{2\chi_1}(L_{\chi_1}-3),
\end{eqnarray*}

\begin{eqnarray*}
g^T_{02q} &=& \frac{1}{4}L_{\Lambda}-\frac{\chi_1}{4(t+\chi_1)}L_{\chi_1}
 -\frac{t}{4(t+\chi_1)}L_t+\frac{3}{8}\, \\
b^T_{02q} &=& \frac{3\chi_1^2-4t\chi_1-t^2}{2(t+\chi_1)^3}L_{\chi_1}+
\frac{t(t+4\chi_1)}{(t+\chi_1)^3}L_t+\frac{t-\chi_1}{2(t+\chi_1)^2}+
\frac{\chi_1^2}{(t+\chi_1)^2}J_{02q}, \\
c^T_{02q} &=& \frac{L_t-L_{\chi_1}}{2(t+\chi_1)}, \quad
\gamma^T_{02q} = \frac{t+2\chi_1}{2(t+\chi_1)^2}
(L_{\chi_1}-L_t)-\frac{1}{2(t+\chi_1)}, \\
a^T_{02q} &=& \alpha^T_{02q}=\beta^T_{02q}=0,
\end{eqnarray*}

\begin{eqnarray}
g^T_{12q} &=& \frac{1}{4}L_{\Lambda} + \frac{t_1L_{t_1}-tL_t}{4(t-t_1)}
+ \frac{3}{8}\, , \nonumber \\
a^T_{12q} &=& \frac{3t^2+4tt_1-t_1^2}{2(t-t_1)^3}L_{t_1}
- \frac{3t^2}{(t-t_1)^3}L_t + \frac{4t-t_1}{(t-t_1)^2}
+ \frac{t^2}{(t-t_1)^2}J_{12q}, \nonumber \\
b^T_{12q} &=& \frac{-t^2+4tt_1+3t_1^2}{2(t-t_1)^3}L_{t_1}
+ \frac{t(t-4t_1)}{(t-t_1)^3}L_t + \frac{3t_1}{(t-t_1)^2}
+ \frac{t_1^2}{(t-t_1)^2}J_{12q}, \nonumber \\
c^T_{12q}&=& \frac{3t_1^2}{(t-t_1)^3}L_{t_1}+
\frac{t^2-4tt_1-3t_1^2}{2(t-t_1)^3}L_t+\frac{4t_1-t}{(t-t_1)^2}+
\frac{t_1^2}{(t-t_1)^2}J_{12q},  \\
\alpha^T_{12q} &=& -\frac{t^2+4tt_1+t_1^2}{2(t-t_1)^3}L_{t_1}+
\frac{t(t+2t_1)}{(t-t_1)^3}L_t-\frac{2t+t_1}{(t-t_1)^2}-
\frac{tt_1}{(t-t_1)^2}J_{12q}, \nonumber \\
\beta^T_{12q}&=&\frac{t_1(5t+t_1)}{2(t-t_1)^3}L_{t_1}-
\frac{t(t+5t_1)}{2(t-t_1)^3}L_t+\frac{3(t+t_1)}{2(t-t_1)^2}+
\frac{tt_1}{(t-t_1)^2}J_{12q}, \nonumber \\
\gamma^T_{12q}&=&-\frac{t_1(t+5t_1)}{2(t-t_1)^3}L_{t_1}+
\frac{-t^2+5tt_1+2t_1^2}{2(t-t_1)^3}L_t+\frac{t-7t_1}{2(t-t_1)^2}-
\frac{t_1^2}{(t-t_1)^2}J_{12q}.\nonumber
\end{eqnarray}

Four-denominator scalar integral reads:
\begin{equation}
J_{012q}=-\frac{1}{t_1\chi_1}\biggl[-L_{\lambda}L_{t_1}+2L_{t_1}L_{\chi_1}
- L_t^2 - 2\Li\biggl(1-\frac{t}{t_1}\biggr) - \frac{\pi^2}{6} \biggr].
\end{equation}

Vector 4-denominator integrals are:
\begin{eqnarray}
a_{012q}&=&\frac{1}{d}\biggl[ - (t\chi_1'+t_1\chi_1)J_{12q}
+ (t+\chi_1)^2J_{02q}-\chi_1(\chi_1'-t_1)J_{01q}-t_1(t+\chi_1)Y\biggr], \nonumber \\
b_{012q}&=&\frac{1}{d}\biggl[ (t_1\chi_1'+t\chi_1)J_{12q}
- (tt_1+\chi_1'\chi_1)J_{02q} + \chi_1(\chi_1-t_1)J_{01q}
+ t_1(t_1-\chi_1)Y\biggr], \nonumber \\
c_{012q}&=&\frac{1}{d}\biggl[ - t_1(\chi_1'+\chi_1)J_{12q}
+ t_1(t+\chi_1)J_{02q} + \chi_1t_1J_{01q}-t_1^2Y\biggr].
\end{eqnarray}
\begin{equation}
Y=J_{012}+\chi_1J_{012q},d=-2t_1\chi_1\chi_1'.
\end{equation}
2-rank 4-denominator tensors are:
\begin{eqnarray}
g^T_{012q} &=& \frac{1}{2}(J_{12q} - \chi_1 c_{012q} ), \nonumber \\
a^{T}_{012q} &=& \frac{1}{d}\biggl[ (t+\chi_1)^2(J_{12q} - \chi_1c_{012q})
- (\chi_1t_1+\chi_1't)a_{12q} + \chi_1(t_1-\chi_1')a_{01q}  \nonumber \\
&-& t_1(t+\chi_1)(a_{012}+\chi_1a_{012q})\biggr], \nonumber \\
b^{T}_{012q} &=& \frac{1}{d}\biggl[ (t_1-\chi_1)^2(J_{12q}-\chi_1c_{012q})
+ (\chi_1't_1+\chi_1t)b_{12q} + \chi_1(\chi_1-t_1)b_{01q} \nonumber \\
&-&(t_1t+\chi_1\chi_1')b_{02q} + t_1(t_1-\chi_1)(a_{012}+\chi_1b_{012q})
\biggr], \nonumber \\
\gamma^{T}_{012q} &=& \frac{1}{d}\biggl[ - t_1(t_1-\chi_1)(J_{12q}
- 2\chi_1c_{012q}) + (\chi_1't_1+\chi_1t)c_{12q} - (tt_1+\chi_1\chi_1')c_{02q}
 \nonumber \\
&+& \chi_1(t_1-\chi_1)b_{01q} \biggr], \nonumber \\
\alpha^{T}_{012q} &=& \frac{1}{d}\biggl[ - (tt_1+\chi_1\chi_1')(J_{12q}
- \chi_1c_{012q}) + (\chi_1't_1+\chi_1t)a_{12q} + \chi_1(\chi_1-t_1)a_{01q}
\nonumber \\
&+& t_1(t_1-\chi_1)(a_{012}+\chi_1a_{012q}) \biggr], \nonumber \\
\beta^{T}_{012q} &=& \frac{1}{d}\biggl[ t_1(t_1+\chi_1')(J_{12q}
- 2\chi_1c_{012q}) - (\chi_1t_1+\chi_1't)c_{12q} + (\chi_1+t)^2c_{02q} \nonumber \\
&+& \chi_1(\chi_1'-t_1)b_{01q} \biggr], \nonumber \\
c^{T}_{012q} &=& \frac{1}{t} \biggl[ J_{12q} - 4g^T_{012q} + t_1\alpha^T_{012q}
+ (\chi_1'-t_1)\beta^T_{012q} + t\gamma^T_{012q}\biggr].
\end{eqnarray}
We put now the coefficients of 3-rank tensor structures:
\begin{eqnarray}
K_{1g} &=& \frac{1}{d}[-(t+\chi_1)^2A_1-t_1(t+\chi_1)A_8
+ (tt_1+\chi_1\chi_1')A_{18}], \nonumber \\
K_{2g} &=& \frac{1}{d}[(tt_1+\chi_1\chi_1')A_1+t_1(t_1-\chi_1)A_8-
(t_1-\chi_1)^2A_{18}], \nonumber \\
K_{qg} &=& \frac{1}{d}[-t_1(t+\chi_1)A_1-t_1^2A_8
+t_1(t_1-\chi_1)A_{18}], \nonumber \\
K_{111} &=& \frac{1}{d}[-(t+\chi_1)^2A_2-t_1(t+\chi_1)A_9
+(tt_1+\chi_1\chi_1')A_{19}], \nonumber \\
K_{112} &=& \frac{1}{d}[(tt_1+\chi_1\chi_1')A_2+t_1(t_1-\chi_1)A_9-
(t_1-\chi_1)^2A_{19}], \nonumber \\
K_{11q} &=& \frac{1}{d}[-t_1(t+\chi_1)A_2-t_1^2A_9
+t_1(t_1-\chi_1)A_{19}], \nonumber \\
K_{12q} &=& \frac{1}{t+\chi_1}[t_1K_{112}+\alpha^T_{12q}
-\alpha^T_{01q}-2K_{1g}], \nonumber \\
K_{1qq} &=& \frac{1}{t+\chi_1}[t_1K_{11q}+\beta^T_{12q}
-\beta^T_{01q}], \nonumber \\
K_{qqq} &=& \frac{1}{t+\chi_1}[t_1K_{1qq}+c^T_{12q}
-c^T_{01q}], \nonumber \\
K_{122} &=& -\frac{1}{t_1}[(t_1-\chi_1)K_{12q}+\alpha^T_{12q}
-\alpha^T_{01q}-2K_{2g}], \nonumber \\
K_{2qq} &=& -\frac{1}{t_1}[(t_1-\chi_1)K_{qqq}+c^T_{12q}
-c^T_{02q}], \nonumber \\
K_{22q} &=& -\frac{1}{t_1}[(t_1-\chi_1)K_{2qq}+\gamma^T_{12q}
-\gamma^T_{02q}], \nonumber \\
K_{222} &=& -\frac{1}{t_1}[(t_1-\chi_1)K_{22q}+b^T_{12q}
-b^T_{02q}],
\end{eqnarray}
where
\begin{eqnarray}
A_1 &=& g^T_{12q}-g^T_{02q}, \quad A_{18}=g^T_{012}-g^T_{01q}+\chi_1g^T_{012q},
\quad A_8=g^T_{12q}-g^T_{01q}, \nonumber \\
A_2 &=& a^T_{12q}-4K_{1g},\quad A_{19}=a^T_{012}-a^T_{01q}+\chi_1a^T_{012q},
\quad A_9=a^T_{12q}-a^T_{01q}.
\end{eqnarray}
We give below some checking equations for coefficients before tensor
structures
of G-type integrals. The complete checking system can be obtained by
contraction
of general tensor expansion with relevant vectors, simplifying the numerators
of the integrand and using a set of vector integrals given above.
Additional check
can be inferred by contraction with metric tensor. In this case the scalar
integrals should be used. The complete set of 10 equations for the
2-rank tensor and 24 equations for the 3-rank 4-denominator tensor integrals
for the G--type was convinced to be fulfilled. For definiteness we give
four equations of such a type, obtained by contraction with metric tensor.
They are:
\begin{eqnarray}
4g^T_{012q}+tc^T_{012q}-t_1\alpha^T_{012q}+(\chi_1-t_1)\beta^T_{012q}+
(t+\chi_1)\gamma^T_{012q} &=& J_{12q}, \nonumber \\
6K_{1g}-t_1K_{112}+(\chi_1-t_1)K_{11q}+tK_{1qq}+(t+\chi_1)K_{12q} &=& a_{12q},
\nonumber \\
6K_{2g}-t_1K_{122}+(\chi_1+t)K_{22q}+tK_{2qq}+(\chi_1-t_1)K_{12q} &=& b_{12q},
\nonumber \\
6K_{qg}+tK_{qqq}+(\chi_1-t_1)K_{1qq}+(t+\chi_1)K_{2qq}-t_1K_{12q} &=& c_{12q}.
\end{eqnarray}
Another indirect check is the absence of infrared divergence containing terms
in all the vector and tensor integrals.

\section*{Appendix B}
{\large \bf Loop integrals for B--type Feynman diagrams}
\vskip 20.0pt
\setcounter{equation}{0}
\renewcommand{\theequation}{B.\arabic{equation}}

We use here the following set of denominators:
\begin{eqnarray}
(1) &=& (p_1-k)^2-m^2,\quad (2)=(p_1-k_1-k)^2-m^2,\quad
(3)=(p_2+k)^2-m^2, \nonumber \\
(4)&=& (p_1-k_1-p_1'-k)^2-\lambda^2, \quad (5)=k^2-\lambda^2.
\end{eqnarray}
4-momentum conservation law we use reads $p_1+p_2=p_1'+p_2'+k_1$.
Scalar products of the loop momentum $k$ with the external 4-vectors
can be expressed in terms of the denominators:
\begin{eqnarray}
&& 2p_1k=(5)-(1),\quad 2p_2k=(3)-(5),\quad 2p_1'k=(4)-(2)-t-\chi_1,
\nonumber \\  \label{B.1}
&& 2k_1k=(2)-(1)+\chi_1,\quad 2p_2'k=(3)-(4)+t.
\end{eqnarray}
Using these relations one can consider only one type of integrals with 5
denominators, namely the scalar one. Using the elegant technique
developed in the paper of Van--Neerven and Vermasseren~\cite{r6}
it can be expressed in the form:
\begin{eqnarray}
J_{12345} &=& -\frac{1}{D}[D_1J_{2345}+D_2J_{1345}+D_3J_{1245}+D_4J_{1235}
+ D_5J_{1234}],\ D=2ss_1t\chi_1\chi_1', \nonumber \\
D_1 &=& s_1t[-t(s-s_1)-s\chi_1-s_1\chi_1'-\chi_1\chi_1'], \nonumber \\
D_2 &=& st[t(s-s_1)+s\chi_1+s_1\chi_1'-\chi_1\chi_1'], \nonumber \\
D_3 &=& \chi_1\chi_1'[-t(s+s_1)-s\chi_1+s_1\chi_1'+\chi_1\chi_1'], \nonumber \\
D_4 &=& s\chi_1[t(s-s_1)+s\chi_1-s_1\chi_1'-\chi_1\chi_1'], \nonumber \\
D_1 &=& s_1\chi_1'[t(s-s_1)-s\chi_1+s_1\chi_1'+\chi_1\chi_1'].
\end{eqnarray}
It is interesting to note that the method described above to calculate the
coefficients of the tensor structures cannot be applied to the
tensor integrals with 5 denominators given above. Some additional
information is needed to close the system of algebraic equations.

We mention a trick which permits to obtain additional equations for vector and
tensor integrals whose denominators do not contain the term $ k^2-\lambda^2$.
It consists in shifting a loop momentum. Thus, for $ J^{\mu}_{1234} $ we
have
$$\int\frac{\dd^4k}{\matr{i}\pi^2}\frac{k}{(1)(2)(3)(4)}
\biggr|_{k=p_1-\tilde k} = \int
\frac{\dd^4\tilde k}{\matr{i}\pi^2}\frac{(p_1-\tilde k)}
{(\tilde 1)(\tilde 2)(\tilde 3)(\tilde 4)}
= p_1J_{1234} +\tilde a(p_1+p_2)+\tilde ck_1+\tilde dp_1' \ , $$
$$(\tilde 1) = \tilde k^2-m^2, \ \ (\tilde 2) = (\tilde k-k_1)^2-m^2 \ , \ \
(\tilde 3) = (p_1+p_2+\tilde k)^2-m^2\ , (\tilde 4) = (\tilde k-p_1'-k_1)^2\ .$$
The comparison of right hand side of this equation with the standard expansion
$$ J^{\mu}_{1234} = (ap_1+bp_2+ck_1+dp_1')^{\mu}_{1234} $$
leads to the new relation:
$$a_{1234} = J_{1234}=b_{1234} \ . $$
Analogous useful relations can be obtained for tensor integrals as well.
We put below the relevant scalar, vector and tensor integrals with 3 and 4
denominators from (B.1) and introduce the parametrization:
\begin{eqnarray}
J_{ij\dots} &=& \int\frac{\dd^4 k}{\matr{i}\pi^2}\frac{1}{(i)(j)\dots}, \quad
J_{ij\dots}^{\mu}=\int\frac{\dd^4 k}{\matr{i}\pi^2}\frac{k^{\mu}}
{(i)(j)\dots}=(a_{ij...}p_1+b_{ij...}p_2 \nonumber \\
&+&c_{ij...}k_1+d_{ij\dots}p_1')^{\mu}, \nonumber \\
J_{ij\dots}^{\mu\nu} &=& \int\frac{\dd^4 k}{\matr{i}\pi^2}
\frac{k^{\mu}k^{\nu}}{(i)(j)\dots}=
(g^Tg+a^Tp_1p_1+b^Tp_2p_2+c^Tk_1k_1+d^Tp_1'p_1'  \\
&+&\alpha^T(p_1p_2)+ \beta^T(p_1k_1)+\gamma^T(p_1p_1')+\rho^T(p_1'p_2)+
\sigma^T(k_1p_2)+\tau^T(p_1'k_1))_{ij\dots}^{\mu\nu}.   \nonumber
\end{eqnarray}
Vector 3-denominator integrals are:
\begin{eqnarray*}
&&a_{245}=-c_{245}=J_{245}+\frac{L_{\chi_1}-L_t}{t+\chi_1},\ b_{245}=0,\\
&&d_{245}=-\frac{\chi_1}{t+\chi_1}J_{245}-\frac{2\chi_1L_{\chi_1}}{(t+\chi_1)^2}
+\frac{(\chi_1-t_1)L_t}{(t+\chi_1)^2},
\end{eqnarray*}

\begin{eqnarray*}
&&a_{145}=-\frac{t}{\chi_1-t_1}J_{145}+
\frac{2\chi_1'L_{\chi_1'}}{(t_1-\chi_1)^2}-\frac{t+\chi_1'}{(\chi_1
-t_1)^2}L_t,\nonumber \\
&&b_{145}=0,\ c_{145}=d_{145}=\frac{L_t-L_{\chi_1'}}{\chi_1'-t},
\end{eqnarray*}

$$a_{345}=-c_{345}=-d_{345}=\frac{L_t}{t},\ b_{345}=-J_{345}+\frac{2L_t}{t},\ $$

$$a_{125}=J_{125}+\frac{L_{\chi_1}}{\chi_1},\ b_{125}=d_{125}=0,\
c_{125}=\frac{L_{\chi_1}-2}{\chi_1},\ $$

\begin{eqnarray*}
&&a_{235}=-c_{235}=\frac{L_{s_1}-L_{\chi_1}}{s-\chi_2},\ d_{235}=0,\nonumber \\
&&b_{235}=-
\frac{\chi_1}{s-\chi_2}J_{235}-\frac{2\chi_1L_{\chi_1}}{(s-\chi_2)^2}+
\frac{\chi_1-s_1}{(s-\chi_2)^2}L_{s_1},
\end{eqnarray*}

$$a_{135}=-b_{135}=\frac{L_s}{s},\ c_{135}=d_{135}=0,\ $$

$$a_{234}=-c_{234}=J_{234}-\frac{L_{s_1}}{s_1},\ b_{234}=-\frac{L_{s_1}}{s_1},\
d_{234}=-J_{234}+\frac{2L_{s_1}}{s_1}\, ,$$

\begin{eqnarray*}
&&a_{123}=J_{123}+b_{123},\ b_{123}=\frac{L_{s_1}-L_s}{s-s_1},\ d_{123}=0,
\nonumber \\
&&c_{123}=
-\frac{s}{s-s_1}J_{123}-\frac{2}{s-s_1}+\frac{2sL_s}{(s-s_1)^2}-
\frac{(s+s_1)L_{s_1}}{(s-s_1)^2},
\end{eqnarray*}

$$\ a_{124}=J_{124},\ b_{124}=0,\ c_{124}=-J_{124}+\frac{L_
{\chi_1'}-2}{\chi_1'},\ d_{124}=-\frac{L_{\chi_1'}}{\chi_1'},$$
\begin{eqnarray}
&& a_{134}=\frac{s}{s-\chi_1'}J_{134}+\frac{2\chi_1'L_{\chi_1'}-
(s+\chi_1')L_s}{(s-\chi_1')^2}, \quad b_{134}=a_{134}-J_{134},\  \nonumber \\
&& c_{134}=d_{134}=-\frac{s}{s-\chi_1'}J_{134}+\frac{-(\chi_1'+s)L_{\chi_1'}+
2sL_s}{(s-\chi_1')^2}.
\end{eqnarray}
Vector integrals with 4 denominators read:
\begin{eqnarray}
a_{1245} &=& \frac{\Delta_{3a}}{\Delta_3}\, ,\quad b_{1245}=0, \quad
c_{1245}=\frac{\Delta_{3c}}{\Delta_3}\, , \nonumber \\
d_{1245}&=&\frac{\Delta_{3d}}{\Delta_3}\, ,\quad
\Delta_3=2t_1\chi_1\chi_1', \nonumber \\
\Delta_{3a} &=& \chi_1'\bigl[\chi_1(2t_1+\chi_1')J_{1245} + \chi_1'J_{124}
- \chi_1J_{125} - (t+\chi_1)J_{245} + (t_1+\chi_1)J_{145} \bigr],
\nonumber \\
\Delta_{3c} &=& t_1\bigl[-\chi_1\chi_1'J_{1245} + \chi_1'J_{124}
+ \chi_1J_{125} - (t+\chi_1)J_{245} + (t-\chi_1')J_{145} \bigr],\nonumber \\
\Delta_{3d} &=& \chi_1\bigl[-\chi_1\chi_1'J_{1245} - \chi_1'J_{124}
+ \chi_1J_{125} + (\chi_1'-t_1)J_{245} + (t-\chi_1')J_{145} \bigr].
\end{eqnarray}

\begin{eqnarray}
a_{1235} &=& \frac{\Delta_{4a}}{\Delta_4}\, ,\quad
b_{1235}=\frac{\Delta_{4b}}{\Delta_4}, \quad
c_{1235}=\frac{\Delta_{4c}}{\Delta_4}\, , \nonumber \\
d_{1235} &=& 0,\quad
\Delta_4=2s\chi_1\chi_2, \nonumber \\
\Delta_{4a} &=& \chi_2\bigl[s\chi_1J_{1235} - (s-s_1)J_{123}
- (s-\chi_2)J_{235} + \chi_1J_{125} + sJ_{135} \bigr],
\nonumber \\
\Delta_{4b} &=& \chi_1\bigl[s\chi_1J_{1235} + (s-s_1)J_{123}
- (s+\chi_2)J_{235} - \chi_1J_{125} + sJ_{135} \bigr],
\nonumber \\
\Delta_{4c} &=& s\bigl[-s\chi_1J_{1235} + (\chi_2-\chi_1)J_{123}
+ (s-\chi_2)J_{235} + \chi_1J_{125} - sJ_{135} \bigr].
\end{eqnarray}

\begin{eqnarray}
a_{1345} &=& \frac{\Delta_{2a}}{\Delta_2}\, ,\quad
b_{1345}=\frac{\Delta_{2b}}{\Delta_2}, \quad
c_{1345}=d_{1345}=\frac{\Delta_{2c}}{\Delta_2}\,,
\quad \Delta_2 = 2stu, \nonumber \\
\Delta_{2a} &=& - st(s+t)J_{1345} + t(s+t)J_{345}
+ s(s+t)J_{135} + (ut-s\chi_1')J_{145} \nonumber \\
&&\hspace{9cm}+(us-t\chi_1')J_{134},\nonumber \\
\Delta_{2b} &=& - st(s+u)J_{1345} + t(s-u)J_{345}
+ s(s+u)J_{135} - (s+u)^2J_{145}  \nonumber \\
&&\hspace{9cm}+ (u\chi_1'-st)J_{134},\nonumber \\
\Delta_{2c} &=& s\bigl[ stJ_{1345} - tJ_{345} - sJ_{135}
+ (s+u)J_{145}+ (t-u)J_{134}\bigr].
\end{eqnarray}

\begin{eqnarray}
a_{2345} &=& - c_{2345} = \frac{\Delta_{1a}}{\Delta_1}\, ,\quad
b_{2345}=\frac{\Delta_{1b}}{\Delta_1}, \quad
d_{2345}=\frac{\Delta_{1d}}{\Delta_1}\, , \nonumber \\
\Delta_1 &=& - 2s_1u_1t, \nonumber \\
\Delta_{1a} &=& - s_1u_1tJ_{2345} - u_1(t+\chi_1)J_{245}
- u_1s_1J_{234} + u_1(s-\chi_2)J_{235} + tu_1J_{345}, \nonumber \\
\Delta_{1b} &=& - s_1t(t+\chi_1)J_{2345} + (t+\chi_1)^2J_{245}
+ s_1(t+\chi_1)J_{234} + (u_1\chi_1+s_1t)J_{235} \nonumber \\
&&+ t(u_1-s_1)J_{345}, \nonumber \\
\Delta_{1c} &=& - s_1t(s-\chi_2)J_{2345} + (u_1\chi_1+s_1t)J_{245}
+ s_1(u_1-t)J_{234} + (s-\chi_2)^2J_{235} \nonumber \\
&&+ t(s-\chi_2)J_{345}.
\end{eqnarray}

\begin{eqnarray}
a_{1234} &=& J_{1234} + \frac{\Delta_{5b}}{\Delta_{5}}, \quad
b_{1234} = \frac{\Delta_{5b}}{\Delta_{5}}, \quad
c_{1234} = - J_{1234} - \frac{\Delta_{5b}}{\Delta_{5}}
+ \frac{\Delta_{5c}}{\Delta_{5}}, \\
d_{1234} &=& - J_{1234} + \frac{\Delta_{5a}}{\Delta_{5}}
- \frac{\Delta_{5b}}{\Delta_{5}},\qquad
\Delta_{5} = 2s_1\chi_1'\chi_2', \quad \chi_2'=s-s_1-\chi_1', \nonumber \\
\Delta_{5a} &=& \chi_2'\bigl[ -(s-s_1)J_{123} + (s-\chi_1')J_{134}
+ \chi_1'J_{124} - s_1J_{234} + s_1\chi_1'J_{1234} \bigr], \nonumber \\
\Delta_{5b} &=& \chi_1'\bigl[ (s-s_1)J_{123} + (2s_1-s+\chi_1')J_{134}
- \chi_1'J_{124} - s_1J_{234} + s_1\chi_1'J_{1234} \bigr], \nonumber \\
\Delta_{5c} &=& s_1\bigl[ (\chi_2'-\chi_1')J_{123} - (s-\chi_1')J_{134}
+ \chi_1'J_{124} + s_1J_{234} - s_1\chi_1'J_{1234} \bigr].\nonumber
\end{eqnarray}

We put now the tensor coefficients for B-type integrals with 4 denominators.
$$g^T_{1245} = \frac{1}{2}\left[2J_{124}-a_{124}-\chi_1 c_{1245} +(t+\chi_1)
d_{1245}\right], $$
\begin{eqnarray*}
a^T_{1245}& =& \frac{1}{t_1\chi_1}\bigl[\chi_1'(-J_{124}+a_{124}-c_{145})
+t_1a_{145} - (t+\chi_1)a_{245} \nonumber \\
&+& t_1\chi_1a_{1245}-\chi_1'(t+\chi_1)d_{1245}\bigr],
\end{eqnarray*}

$$c^T_{1245} = \frac{1}{\chi_1\chi_1'}\left[t_1(-J_{124}+a_{124})+\chi_1c_{125}
+(t_1-\chi_1)c_{145} - \chi_1\chi_1'c_{1245}\right] \ , $$
$$d^T_{1245} = \frac{1}{t_1\chi_1'}\left[\chi_1(-J_{124}+a_{124}-a_{245})
+(t_1-\chi_1)c_{145} -t_1d_{245}- \chi_1\chi_1'd_{1245}\right] \ , $$
$$\beta^T_{1245} = \frac{1}{\chi_1}\left[-J_{124}+a_{124}+c_{145}
+\chi_1 c_{1245}\right], $$
$$\gamma^T_{1245} = \frac{1}{t_1}\left[J_{124}-a_{124}+a_{245}+c_{145}+
(t+\chi_1)d_{1245}\right],$$
$$\tau^T_{1245} = \frac{1}{\chi_1'}\left[-J_{124}+a_{245}+\chi_1c_{1245}-
(t+\chi_1)d_{1245}\right],$$
\begin{equation}\label{I1}
b^T_{1245}=\alpha^T_{1245}=\rho^T_{1245}=\sigma^T_{1245}=0 \ .
\end{equation}
As a check one can use the result of contraction by the metric tensor:
\begin{equation}
4g^T_{1245} + \chi_1\beta^T_{1245} - t_1\gamma^T_{1245}
+ \chi_1'\tau^T_{1245} = J_{124}.
\end{equation}

$$g^T_{1235} = \frac{1}{2}[2J_{123}-a_{123}+b_{123}-\chi_1 c_{1235}] \ , $$
$$a^T_{1235} = \frac{1}{s\chi_1}[\chi_2J_{123}-(\chi_1+\chi_2)a_{123}
+\chi_1a_{125})-\chi_1\chi_2c_{1235}], $$
$$b^T_{1235} = \frac{1}{s\chi_2}[\chi_1(J_{123}-a_{235})+(\chi_1+\chi_2)b_{123}
-\chi_2b_{235} - \chi_1^2c_{1235}] \ , $$
$$c^T_{1235} = \frac{1}{\chi_1\chi_2}[s(J_{123}+b_{123})
-(s-\chi_2)a_{235} + \chi_2c_{123}-s\chi_1c_{1235}] \ , $$
$$\alpha^T_{1235} = \frac{1}{s}[-J_{123}+a_{123}-a_{235}-b_{123}],$$
$$\beta^T_{1235} = \frac{1}{\chi_1}[-J_{123}+a_{123}+\chi_1 c_{1235}],$$
$$\sigma^T_{1235} = \frac{1}{\chi_2}\bigl[-J_{123}+a_{235}-b_{123}+
\chi_1c_{1235} \bigr], $$
\begin{equation}\label{I2}
d^T_{1235}=\gamma^T_{1235}=\rho^T_{1235}=\tau^T_{1235}=0 \ .
\end{equation}
One of the checking relations here has the form
\begin{equation}
4g^T_{1235} + s\alpha^T_{1235} + \chi_1\beta^T_{1235}
+ \chi_2\sigma^T_{1235} = J_{123}.
\end{equation}

\begin{eqnarray}
g^T_{1345} &=&\frac{1}{2}[J_{134}+tc_{1345}], \nonumber \\
a^T_{1345} &=& \frac{1}{st(\chi_1'-s-t)}[(s+t)^2J_{134}+t(\chi_1'-s-t)a_{145}-
(s(s+t)+t\chi_1')a_{134} \nonumber \\
&+& \chi_1'(s+t)(c_{145}-c_{134})+t(s+t)^2c_{1345}], \nonumber \\
b^T_{1345} &=& \frac{1}{s}[b_{134}-b_{345}-(\chi_1'-t)\rho^T_{1345}], \nonumber \\
c^T_{1345} &=& d^T_{1345}=\tau^T_{1345}=\frac{1}{t(\chi_1'-s-t)}[(\chi_1'-t)
(c_{145}-c_{134})-s(b_{134}-tc_{1345})], \nonumber \\
\alpha^T_{1345} &=& \frac{1}{st(\chi_1'-s-t)}[-t(\chi_1'-s-t)a_{345}+\chi_1'
(\chi_1'-t)(c_{145}-c_{134}) \nonumber \\
&-& s\chi_1'(a_{134}-J_{134}) + st\chi_1'c_{1345}], \nonumber \\
\beta^T_{1345} &=& \gamma^T_{1345}=\frac{1}{t(\chi_1'-s-t)}[(s+t)
(b_{134}-tc_{1345})-\chi_1'(c_{145}-c_{134})],
\nonumber \\ \label{I3}
\rho^T_{1345} &=& \sigma^T_{1345} =
\frac{1}{st(\chi_1'-s-t)} [ - (\chi_1'-t)^2c_{145}
+ t(\chi_1'-s-t)a_{345} \nonumber \\
&+&(\chi_1'(\chi_1'-t)-st)c_{134}+s(\chi_1'-t)b_{134}-st(\chi_1'-t)c_{1345}].
\end{eqnarray}
The relation of the same type for the above coefficients reads:
\begin{equation}
4g^T_{1345} + \chi_1'c^T_{1345} + s\alpha^T_{1345}
+ (\chi_1-t_1)\beta^T_{1345}
+ (\chi_2-u_1)\sigma^T_{1345} = J_{134}.
\end{equation}

\begin{eqnarray}
g^T_{2345} &=& \frac{1}{2}\bigl[ J_{234} + \chi_1a_{2345}
+ (t+\chi_1)d_{1345}\bigr],\nonumber  \\
a^T_{2345} &=& c^T_{2345}=-\beta^T_{2345}=\frac{1}{s_1t}\bigl[ - ta_{345}
- (s_1+\chi_1)a_{235}+s_1ta_{2345}\bigr],\nonumber \\
b^T_{2345} &=& \frac{1}{s_1t(\chi_1+s_1+t)}\bigl[ s_1t(b_{235}-b_{345})
- \chi_1(\chi_1+t)a_{235} -t(t+\chi_1)a_{345}\nonumber \\
&-& s_1t(\chi_1+t)b_{2345} \bigr],\nonumber \\
d^T_{2345} &=& \frac{1}{\chi_1+s_1+t}\biggl[ d_{245} - d_{234}
- \frac{\chi_1+s_1}{s_1t(\chi_1+s_1+t)}\biggl( s_1t(a_{245}-a_{234})\nonumber \\
&+& t(\chi_1+s_1)a_{345}
+(\chi_1+s_1)^2a_{235}-s_1t(\chi_1+s_1)a_{2345} \biggr)\biggr],\nonumber \\
\alpha^T_{2345} &=& -\sigma^T_{2345}=\frac{1}{s_1t}\bigl[ - \chi_1a_{235}
-ta_{345}\bigr],\nonumber \\
\gamma^T_{2345} &=& -\tau^T_{2345}=\frac{1}{s_1t(\chi_1+s_1+t)}[s_1t(a_{245}
-a_{234}) + t(\chi_1+s_1)a_{345} \nonumber \\
&+& (\chi_1+s_1)^2a_{235}- s_1t(\chi_1+s_1)a_{2345}] \nonumber \\
\rho^T_{2345} &=& \frac{1}{s_1t(\chi_1+s_1+t)}[-s_1ta_{234}+\chi_1(\chi_1+s_1)
a_{235}+t(\chi_1+s_1)a_{345}  \nonumber \\
&-& s_1t\chi_1a_{2345}-s_1t(\chi_1+t)d_{2345}].
\end{eqnarray}
The above coefficients have to satisfy the relation
\begin{eqnarray*}
4g^T_{2345} - \chi_1a^T_{2345} + (s-\chi_2)\alpha^T_{2345}
- (t+\chi_1)\gamma^T_{2345}
- u_1\rho^T_{2345} = J_{234}.
\end{eqnarray*}

\begin{eqnarray}
&& g^T_{1234} = \frac{1}{2}\Biggl[J_{123} - \chi_1'\frac{\Delta^{(3)}}{\Delta}
\Biggl] ,  \nonumber  \\
&& a^T_{1234} = 2\frac{\Delta^{(2)}}{\Delta} + J_{1234} + \tilde b_{1234}\ ,
\nonumber \\
&& b^T_{1234} = \tilde b_{1234} \ ,  \nonumber  \\
&& c^T_{1234} = 2\frac{\Delta^{(2)}}{\Delta} - 2\frac{\Delta^{(3)}}{\Delta}
+ J_{1234} + \tilde b_{1234} + \tilde c_{1234} - 2\tilde\gamma_{1234}\ ,
\nonumber \\
&& d^T_{1234} = 2\frac{\Delta^{(2)}}{\Delta} - 2\frac{\Delta^{(1)}}{\Delta}
+ J_{1234} + \tilde b_{1234} + \tilde a_{1234} - 2\tilde\alpha_{1234}\ ,
\nonumber  \\
&& \alpha^T_{1234} = \frac{\Delta^{(2)}}{\Delta} + \tilde b_{1234}\ ,
\nonumber \\
&& \beta^T_{1234} = \frac{\Delta^{(3)}}{\Delta} - 2\frac{\Delta^{(2)}}{\Delta}
- J_{1234} - \tilde b_{1234} + \tilde\gamma_{1234}\, \nonumber \\
&& \gamma^T_{1234} = \frac{\Delta^{(1)}}{\Delta} - 2\frac{\Delta^{(2)}}{\Delta}
- J_{1234} - \tilde b_{1234} + \tilde\alpha_{1234}\, \nonumber \\
&& \rho^T_{1234} = -\frac{\Delta^{(2)}}{\Delta} - \tilde b_{1234}
+ \tilde\alpha_{1234}\ , \nonumber \\
&& \sigma^T_{1234} = -\frac{\Delta^{(2)}}{\Delta} - \tilde b_{1234}
+\tilde\gamma_{1234}\ , \nonumber \\
&& \tau^T_{1234} = 2\frac{\Delta^{(2)}}{\Delta} - \frac{\Delta^{(1)}}{\Delta}
-\frac{\Delta^{(3)}}{\Delta}+ J_{1234} + \tilde b_{1234}
+\tilde\beta_{1234}-\tilde\alpha_{1234}-\tilde\gamma_{1234},
\end{eqnarray}
where the quantities with the sign $\tilde{}$ are defined as follows:
\begin{eqnarray}
&&\tilde a_{1234} = \frac{1}{s\chi_1'}(L_s-L_{s_1}-L_{\chi_1'}) \ , \nonumber \\
&&\tilde b_{1234} = \frac{1}{\chi_2'}\left[\chi_1'\frac{\Delta^{(2)}}
{\Delta} + \frac{\chi_1'}{s-\chi_1'}J_{134} + \frac{L_{s1}}{s_1}-\frac{L_s}
{s-\chi_1'} + \frac{\chi_1'(s_1-\chi_2')}{s_1(s-\chi_1')^2}(L_{\chi_1'}-L_s)
\right] \ , \nonumber \\
&&\tilde c_{1234} = \frac{1}{\chi_2'}\Biggl[\frac{s_1^2}{\chi_2'}\frac{\Delta^
{(2)}}{\Delta} - \frac{s_1}{\chi_2'}J_{124} + \bigl(\frac{s_1}{\chi_2'}+\frac
{s_1}{s-s_1}\bigr)J_{123} + \frac{2-L_s}{s-s_1}  \nonumber \\
&& - \frac{2-L_{\chi_1'}}{\chi_1'} - \frac{2s_1}{(s-s_1)^2}(L_s-L_{s_1})
\Biggr] \ , \nonumber \\
&&\tilde\alpha_{1234} = \frac{L_{\chi_1'}-L_s}{s_1(s-\chi_1')}\, , \qquad
\tilde\beta_{1234} = \frac{\Delta^{(3)}}{\Delta} - \frac{L_s-L_{s_1}}
{\chi_1'(s-s_1)} \, , \nonumber \\
&&\tilde\gamma_{1234} = \frac{1}{\chi_2'}\left[\chi_1'\frac{\Delta^{(3)}}
{\Delta} - J_{123} + \frac{L_s-L_{s_1}}{s-s_1} +
\frac{L_s-L_{\chi_1'}}{s-\chi_1'}\right].
\end{eqnarray}
One of the checking relations takes the form
\begin{eqnarray*}
2g^T_{1234} + \chi_1\beta^T_{1234} + \chi_2\sigma^T_{1234}
+ \chi_1'\tau^T_{1234} = a_{134} - a_{234} +\chi_1a_{1234}.
\end{eqnarray*}

At the end of this Appendix we give the table of scalar integrals with two,
three and four denominators. We imply the real part everywhere
and the ultraviolet asymptotic is assumed as well.
\begin{eqnarray}
&& J_{12} = -1 + L_{\Lambda} \ , \qquad
J_{13} = 1 + L_{\Lambda} - L_s  \ , \nonumber \\
&& J_{14} = 1 + L_{\Lambda} - L_{\chi_1'}  \ , \qquad
J_{15}=J_{24}=J_{34}=J_{35}=L_{\Lambda}+1 \ ,\nonumber  \\
&& J_{23} = 1 + L_{\Lambda} - L_{s_1}  \ , \qquad
J_{25} = 1 + L_{\Lambda} - L_{\chi_1}  \ , \nonumber \\
&& J_{45} = 1 + L_{\Lambda} - L_t,
\end{eqnarray}
where
\begin{eqnarray}
&& L_{\Lambda}=\ln\frac{\Lambda^2}{m^2}, \quad
L_s=\ln\frac{s}{m^2}, \quad
L_{\lambda}=\ln\frac{\lambda^2}{m^2}, \nonumber \\
&& L_{s_1}=\ln\frac{s_1}{m^2}, \quad
L_{\chi_1'}=\ln\frac{\chi_1'}{m^2}, \quad
L_{\chi_1}=\ln\frac{\chi_1}{m^2}, \quad
L_t=\ln\frac{-t}{m^2}.
\end{eqnarray}

3-denominator scalar integrals are
\begin{eqnarray}
J_{123} &=& \frac{1}{2(s-s_1)}(L_s^2-L_{s_1}^2), \quad
J_{345} = \frac{1}{t}\left[\frac{1}{2}L_t^2+\frac{2\pi^2}{3}\right],
\nonumber \\
J_{124} &=& \frac{1}{\chi_1'}\left[\frac{1}{2}L_{\chi_1'}^2
-\frac{\pi^2}{6}\right], \quad
J_{125} = \frac{1}{\chi_1}\left[-\frac{1}{2}L_{\chi_1}^2
-\frac{\pi^2}{3}\right], \nonumber \\
J_{134} &=& \frac{1}{s-\chi_1'}\left[\frac{3}{2}L_s^2+\frac{1}{2}L_{\chi_1'}^2-
2L_sL_{\chi_1'}+2\Li\left(1-\frac{\chi_1'}{s}\right)\right], \nonumber \\
J_{235} &=& \frac{1}{s_1+\chi_1}\left[\frac{3}{2}L_{s_1}^2
+\frac{1}{2}L_{\chi_1}^2-
2L_{s_1}L_{\chi_1}+2\Li\left(1+\frac{\chi_1}{s_1}\right)
-\frac{3\pi^2}{2}\right], \nonumber \\
J_{135} &=& \frac{1}{s}\left[\frac{1}{2}L_s^2-
L_sL_{\lambda}-\frac{2\pi^2}{3}\right], \
J_{234} = \frac{1}{s_1}\left[\frac{1}{2}L_{s_1}^2-
L_{s_1}L_{\lambda}-\frac{2\pi^2}{3}\right], \nonumber \\
J_{245} &=& \frac{1}{t+\chi_1}\left[\frac{1}{2}L_t^2-\frac{1}{2}L_{\chi_1}^2+
2\Li\left(1+\frac{\chi_1}{t}\right)\right], \nonumber \\
J_{145} &=& \frac{1}{-t+\chi_1'}\left[\frac{1}{2}L_{\chi_1'}^2-
\frac{1}{2}L_t^2-\frac{\pi^2}{2}
-2\Li\left(1-\frac{\chi_1'}{t}\right)\right].
\end{eqnarray}
4-denominator scalar integrals read:
\begin{eqnarray}
&&J_{1245}=\frac{1}{\chi_1\chi_1'}\left[-L_{\chi_1}^2-L_{\chi_1'}^2-
L_t^2-2L_{\chi_1}L_{\chi_1'}+2L_{\chi_1}L_t+2L_{\chi_1'}L_t
+\frac{2\pi^2}{3}\right], \nonumber \\
&&J_{2345}=\frac{1}{s_1t}\left[L_{s_1}^2-L_{s_1}L_{\lambda}-
2L_{s_1}L_{\chi_1}+2L_{s_1}L_t-\frac{5\pi^2}{6}\right] , \nonumber \\
&&J_{1345}=\frac{1}{st}\left[L_s^2-L_sL_{\lambda}-
2L_sL_{\chi_1'}+2L_sL_t+7\frac{\pi^2}{6}\right] , \\
&&J_{1235}=\frac{1}{s\chi_1}\left[L_{s_1}^2+L_sL_{\lambda}-
2L_sL_{\chi_1}+2\Li\left(1-\frac{s_1}{s}\right)-5\frac{\pi^2}{6}\right],
 \nonumber \\
&&J_{1234}=\frac{1}{s_1\chi_1'}\left[-L_s^2-L_{s_1}L_{\lambda}+
2L_{s_1}L_{\chi_1'}-2\Li\left(1-\frac{s}{s_1}\right)-7\frac{\pi^2}{6}\right].
\nonumber
\end{eqnarray}

\newpage

\unitlength=0.55mm
\special{em:linewidth 0.6pt}
\linethickness{0.6pt}
\begin{picture}(283.81,269.00)
\put(37.00,225.00){\oval(3.00,3.00)[l]}
\put(37.00,228.00){\oval(3.00,3.00)[r]}
\put(37.00,231.00){\oval(3.00,3.00)[l]}
\put(37.00,234.00){\oval(3.00,3.00)[r]}
\put(37.00,237.00){\oval(3.00,3.00)[l]}
\put(14.00,269.00){\line(4,-5){20.00}}
\put(40.00,244.00){\line(4,5){20.00}}
\put(37.14,223.00){\line(-1,-1){20.00}}
\put(37.14,223.00){\line(1,-1){20.00}}
\put(49.00,238.00){\makebox(0,0)[cc]{$(1)$}}
\put(42.81,229.00){\makebox(0,0)[cc]{$\gamma^{*}$}}
\put(12.00,264.00){\makebox(0,0)[cc]{$e^-$}}
\put(61.00,264.00){\makebox(0,0)[cc]{$e^-$}}
\put(12.00,209.00){\makebox(0,0)[cc]{$e^+$}}
\put(61.00,209.00){\makebox(0,0)[cc]{$e^+$}}
\put(37.00,242.00){\circle{6.00}}
\put(26.00,253.50){\vector(1,-1){0.97}}
\put(47.00,253.00){\vector(1,1){0.97}}
\put(46.00,214.00){\vector(-1,1){0.92}}
\put(30.00,216.00){\vector(-1,-1){1.06}}
\put(37.00,247.00){\oval(3.00,3.00)[l]}
\put(37.00,250.00){\oval(3.00,3.00)[r]}
\put(37.00,253.00){\oval(3.00,3.00)[l]}
\put(37.00,256.00){\oval(3.00,3.00)[r]}
\put(37.00,259.00){\oval(3.00,3.00)[l]}
\put(37.00,262.00){\oval(3.00,3.00)[r]}
\
\put(107.00,229.00){\oval(3.00,3.00)[l]}
\put(107.00,232.00){\oval(3.00,3.00)[r]}
\put(107.00,235.00){\oval(3.00,3.00)[l]}
\put(107.00,238.00){\oval(3.00,3.00)[r]}
\put(107.00,241.00){\oval(3.00,3.00)[l]}
\put(107.00,204.00){\oval(3.00,3.00)[l]}
\put(107.00,207.00){\oval(3.00,3.00)[r]}
\put(107.00,210.00){\oval(3.00,3.00)[l]}
\put(107.00,213.00){\oval(3.00,3.00)[r]}
\put(107.00,216.00){\oval(3.00,3.00)[l]}
\put(107.00,219.00){\oval(3.00,3.00)[r]}
\put(87.00,268.00){\line(4,-5){20.00}}
\put(107.00,243.00){\line(4,5){20.00}}
\put(104.14,222.00){\line(-1,-1){20.00}}
\put(110.14,222.00){\line(1,-1){20.00}}
\put(127.14,205.00){\vector(-1,1){9.00}}
\put(97.00,255.50){\vector(3,-4){1.67}}
\put(116.00,254.25){\vector(3,4){1.67}}
\put(119.00,238.00){\makebox(0,0)[cc]{$(2)$}}
\put(82.00,264.00){\makebox(0,0)[cc]{$p_1$}}
\put(131.00,264.00){\makebox(0,0)[cc]{$p_1'$}}
\put(83.00,211.00){\makebox(0,0)[cc]{$-p_2$}}
\put(130.00,211.00){\makebox(0,0)[cc]{$-p_2'$}}
\
\put(166.00,233.00){\oval(3.00,3.00)[b]}
\put(169.00,233.00){\oval(3.00,3.00)[t]}
\put(172.00,233.00){\oval(3.00,3.00)[b]}
\put(175.00,233.00){\oval(3.00,3.00)[t]}
\put(178.00,233.00){\oval(3.00,3.00)[b]}
\put(181.00,233.00){\oval(3.00,3.00)[t]}
\put(202.14,253.00){\line(-1,-1){20.00}}
\put(182.14,233.00){\line(1,-1){20.00}}
\put(202.81,238.00){\makebox(0,0)[cc]{$(3)$}}
\put(218.14,254.00){\line(1,-1){20.00}}
\put(238.14,234.00){\line(-1,-1){20.00}}
\put(278.81,239.00){\makebox(0,0)[cc]{$(4)$}}
\put(107.00,224.00){\circle{6.00}}
\put(96.00,214.00){\vector(-1,-1){1.05}}
\put(161.00,233.00){\circle{6.00}}
\put(159.00,236.00){\line(-1,1){16.93}}
\put(159.00,230.00){\line(-1,-1){16.93}}
\put(151.00,222.00){\vector(-1,-1){0.98}}
\put(151.00,244.00){\vector(1,-1){1.04}}
\put(191.00,242.00){\vector(1,1){1.04}}
\put(193.00,222.00){\vector(-1,1){0.96}}
\put(257.00,235.00){\circle{6.00}}
\put(258.00,232.00){\line(0,0){0.00}}
\put(259.00,237.00){\line(1,1){17.01}}
\put(259.00,233.00){\line(5,-6){15.82}}
\put(227.00,245.00){\vector(1,-1){0.95}}
\put(229.00,225.00){\vector(-1,-1){1.05}}
\put(266.00,224.00){\vector(-1,2){1.00}}
\put(267.00,245.00){\vector(1,1){0.97}}
\put(161.00,238.00){\oval(3.00,3.00)[l]}
\put(161.00,241.00){\oval(3.00,3.00)[r]}
\put(161.00,244.00){\oval(3.00,3.00)[l]}
\put(161.00,247.00){\oval(3.00,3.00)[r]}
\put(161.00,250.00){\oval(3.00,3.00)[l]}
\put(161.00,253.00){\oval(3.00,3.00)[r]}
\put(240.00,234.00){\oval(3.00,3.00)[b]}
\put(243.00,234.00){\oval(3.00,3.00)[t]}
\put(246.00,234.00){\oval(3.00,3.00)[b]}
\put(249.00,234.00){\oval(3.00,3.00)[t]}
\put(252.00,234.00){\oval(3.00,3.00)[b]}
\put(257.00,240.00){\oval(3.00,3.00)[l]}
\put(257.00,243.00){\oval(3.00,3.00)[r]}
\put(257.00,246.00){\oval(3.00,3.00)[l]}
\put(257.00,249.00){\oval(3.00,3.00)[r]}
\put(257.00,252.00){\oval(3.00,3.00)[l]}
\put(257.00,255.00){\oval(3.00,3.00)[r]}
\put(37.00,150.00){\oval(3.00,3.00)[l]}
\put(37.00,153.00){\oval(3.00,3.00)[r]}
\put(37.00,156.00){\oval(3.00,3.00)[l]}
\put(37.00,159.00){\oval(3.00,3.00)[r]}
\put(37.00,162.00){\oval(3.00,3.00)[l]}
\put(14.00,194.00){\line(4,-5){20.00}}
\put(40.00,169.00){\line(4,5){20.00}}
\put(37.14,148.00){\line(-1,-1){20.00}}
\put(37.14,148.00){\line(1,-1){20.00}}
\put(49.00,163.00){\makebox(0,0)[cc]{$(5)$}}
\put(37.00,167.00){\circle*{6.00}}
\put(26.00,178.50){\vector(1,-1){0.97}}
\put(47.00,178.00){\vector(1,1){0.97}}
\put(46.00,139.00){\vector(-1,1){0.92}}
\put(30.00,141.00){\vector(-1,-1){1.06}}
\put(37.00,172.00){\oval(3.00,3.00)[l]}
\put(37.00,175.00){\oval(3.00,3.00)[r]}
\put(37.00,178.00){\oval(3.00,3.00)[l]}
\put(37.00,181.00){\oval(3.00,3.00)[r]}
\put(37.00,184.00){\oval(3.00,3.00)[l]}
\put(37.00,187.00){\oval(3.00,3.00)[r]}
\put(107.00,154.00){\oval(3.00,3.00)[l]}
\put(107.00,157.00){\oval(3.00,3.00)[r]}
\put(107.00,160.00){\oval(3.00,3.00)[l]}
\put(107.00,163.00){\oval(3.00,3.00)[r]}
\put(107.00,166.00){\oval(3.00,3.00)[l]}
\put(107.00,129.00){\oval(3.00,3.00)[l]}
\put(107.00,132.00){\oval(3.00,3.00)[r]}
\put(107.00,135.00){\oval(3.00,3.00)[l]}
\put(107.00,138.00){\oval(3.00,3.00)[r]}
\put(107.00,141.00){\oval(3.00,3.00)[l]}
\put(107.00,144.00){\oval(3.00,3.00)[r]}
\put(87.00,193.00){\line(4,-5){20.00}}
\put(107.00,168.00){\line(4,5){20.00}}
\put(104.14,147.00){\line(-1,-1){20.00}}
\put(110.14,147.00){\line(1,-1){20.00}}
\put(127.14,130.00){\vector(-1,1){9.00}}
\put(97.00,180.50){\vector(3,-4){1.67}}
\put(116.00,179.25){\vector(3,4){1.67}}
\put(119.00,163.00){\makebox(0,0)[cc]{$(6)$}}
\put(166.00,158.00){\oval(3.00,3.00)[b]}
\put(169.00,158.00){\oval(3.00,3.00)[t]}
\put(172.00,158.00){\oval(3.00,3.00)[b]}
\put(175.00,158.00){\oval(3.00,3.00)[t]}
\put(178.00,158.00){\oval(3.00,3.00)[b]}
\put(181.00,158.00){\oval(3.00,3.00)[t]}
\put(202.14,178.00){\line(-1,-1){20.00}}
\put(182.14,158.00){\line(1,-1){20.00}}
\put(202.81,163.00){\makebox(0,0)[cc]{$(7)$}}
\put(218.14,179.00){\line(1,-1){20.00}}
\put(238.14,159.00){\line(-1,-1){20.00}}
\put(278.81,164.00){\makebox(0,0)[cc]{$(8)$}}
\put(107.00,149.00){\circle*{6.00}}
\put(96.00,139.00){\vector(-1,-1){1.05}}
\put(161.00,158.00){\circle*{6.00}}
\put(159.00,161.00){\line(-1,1){16.93}}
\put(159.00,155.00){\line(-1,-1){16.93}}
\put(151.00,147.00){\vector(-1,-1){0.98}}
\put(151.00,169.00){\vector(1,-1){1.04}}
\put(191.00,167.00){\vector(1,1){1.04}}
\put(193.00,147.00){\vector(-1,1){0.96}}
\put(257.00,160.00){\circle*{6.00}}
\put(258.00,157.00){\line(0,0){0.00}}
\put(259.00,162.00){\line(1,1){17.01}}
\put(259.00,158.00){\line(5,-6){15.82}}
\put(227.00,170.00){\vector(1,-1){0.95}}
\put(229.00,150.00){\vector(-1,-1){1.05}}
\put(266.00,149.00){\vector(-1,2){1.00}}
\put(267.00,170.00){\vector(1,1){0.97}}
\put(161.00,163.00){\oval(3.00,3.00)[l]}
\put(161.00,166.00){\oval(3.00,3.00)[r]}
\put(161.00,169.00){\oval(3.00,3.00)[l]}
\put(161.00,172.00){\oval(3.00,3.00)[r]}
\put(161.00,175.00){\oval(3.00,3.00)[l]}
\put(161.00,178.00){\oval(3.00,3.00)[r]}
\put(240.00,159.00){\oval(3.00,3.00)[b]}
\put(243.00,159.00){\oval(3.00,3.00)[t]}
\put(246.00,159.00){\oval(3.00,3.00)[b]}
\put(249.00,159.00){\oval(3.00,3.00)[t]}
\put(252.00,159.00){\oval(3.00,3.00)[b]}
\put(257.00,165.00){\oval(3.00,3.00)[l]}
\put(257.00,168.00){\oval(3.00,3.00)[r]}
\put(257.00,171.00){\oval(3.00,3.00)[l]}
\put(257.00,174.00){\oval(3.00,3.00)[r]}
\put(257.00,177.00){\oval(3.00,3.00)[l]}
\put(257.00,180.00){\oval(3.00,3.00)[r]}
\
\
\put(34.50,100.50){\oval(21.00,7.00)[]}
\put(24.00,100.50){\line(-1,0){12.87}}
\put(58.00,100.50){\line(-1,0){12.87}}
\put(27.00,83.00){\oval(3.00,3.00)[r]}
\put(27.00,86.00){\oval(3.00,3.00)[l]}
\put(27.00,89.00){\oval(3.00,3.00)[r]}
\put(27.00,92.00){\oval(3.00,3.00)[l]}
\put(27.00,95.00){\oval(3.00,3.00)[r]}
\put(42.00,83.00){\oval(3.00,3.00)[r]}
\put(42.00,86.00){\oval(3.00,3.00)[l]}
\put(42.00,89.00){\oval(3.00,3.00)[r]}
\put(42.00,92.00){\oval(3.00,3.00)[l]}
\put(42.00,95.00){\oval(3.00,3.00)[r]}
\put(11.00,81.50){\line(1,0){47.14}}
\put(35.00,105.50){\oval(3.00,3.00)[r]}
\put(35.00,108.50){\oval(3.00,3.00)[l]}
\put(35.00,111.50){\oval(3.00,3.00)[r]}
\put(35.00,114.50){\oval(3.00,3.00)[l]}
\put(35.00,117.50){\oval(3.00,3.00)[r]}
\put(16.00,100.50){\vector(1,0){1.10}}
\put(50.00,100.50){\vector(1,0){1.07}}
\put(52.00,81.50){\vector(-1,0){0.93}}
\put(18.00,81.50){\vector(-1,0){0.90}}
\
\put(84.00,119.00){\line(1,0){47.14}}
\put(90.00,119.00){\vector(1,0){1.97}}
\put(121.00,119.00){\vector(1,0){1.02}}
\put(100.00,105.00){\oval(3.00,3.00)[r]}
\put(100.00,108.00){\oval(3.00,3.00)[l]}
\put(100.00,111.00){\oval(3.00,3.00)[r]}
\put(100.00,114.00){\oval(3.00,3.00)[l]}
\put(100.00,117.00){\oval(3.00,3.00)[r]}
\put(115.00,105.00){\oval(3.00,3.00)[r]}
\put(115.00,108.00){\oval(3.00,3.00)[l]}
\put(115.00,111.00){\oval(3.00,3.00)[r]}
\put(115.00,114.00){\oval(3.00,3.00)[l]}
\put(115.00,117.00){\oval(3.00,3.00)[r]}
\put(107.50,99.50){\oval(21.00,7.00)[]}
\put(97.00,99.50){\line(-1,0){12.87}}
\put(131.00,99.50){\line(-1,0){12.87}}
\put(89.00,99.50){\vector(1,0){1.10}}
\put(123.00,99.50){\vector(1,0){1.07}}
\put(107.00,82.50){\oval(3.00,3.00)[r]}
\put(107.00,85.50){\oval(3.00,3.00)[l]}
\put(107.00,88.50){\oval(3.00,3.00)[r]}
\put(107.00,91.50){\oval(3.00,3.00)[l]}
\put(107.00,94.50){\oval(3.00,3.00)[r]}
\put(165.50,91.00){\oval(7.00,21.00)[]}
\put(163.00,81.00){\line(-1,0){14.91}}
\put(163.00,101.00){\line(-1,0){14.91}}
\put(166.00,103.50){\oval(3.00,3.00)[r]}
\put(166.00,106.50){\oval(3.00,3.00)[l]}
\put(166.00,109.50){\oval(3.00,3.00)[r]}
\put(166.00,112.50){\oval(3.00,3.00)[l]}
\put(166.00,115.50){\oval(3.00,3.00)[r]}
\put(170.50,100.50){\oval(3.00,3.00)[t]}
\put(173.50,100.50){\oval(3.00,3.00)[b]}
\put(176.50,100.50){\oval(3.00,3.00)[t]}
\put(179.50,100.50){\oval(3.00,3.00)[b]}
\put(182.50,100.50){\oval(3.00,3.00)[t]}
\put(170.50,82.50){\oval(3.00,3.00)[t]}
\put(173.50,82.50){\oval(3.00,3.00)[b]}
\put(176.50,82.50){\oval(3.00,3.00)[t]}
\put(179.50,82.50){\oval(3.00,3.00)[b]}
\put(182.50,82.50){\oval(3.00,3.00)[t]}
\put(184.10,101.00){\line(0,-1){20.00}}
\put(199.00,101.00){\line(-1,0){14.91}}
\put(199.00,81.00){\line(-1,0){14.91}}
\put(153.00,101.00){\vector(1,0){0.98}}
\put(154.00,81.00){\vector(-1,0){1.01}}
\put(190.00,101.00){\vector(1,0){1.05}}
\put(191.00,81.00){\vector(-1,0){0.94}}
\
\put(257.00,91.00){\oval(7.00,21.00)[]}
\put(239.00,81.00){\line(-1,0){14.91}}
\put(239.00,101.00){\line(-1,0){14.91}}
\put(257.00,103.50){\oval(3.00,3.00)[r]}
\put(257.00,106.50){\oval(3.00,3.00)[l]}
\put(257.00,109.50){\oval(3.00,3.00)[r]}
\put(257.00,112.50){\oval(3.00,3.00)[l]}
\put(257.00,115.50){\oval(3.00,3.00)[r]}
\put(240.50,100.50){\oval(3.00,3.00)[t]}
\put(243.50,100.50){\oval(3.00,3.00)[b]}
\put(246.50,100.50){\oval(3.00,3.00)[t]}
\put(249.50,100.50){\oval(3.00,3.00)[b]}
\put(252.50,100.50){\oval(3.00,3.00)[t]}
\put(240.50,81.50){\oval(3.00,3.00)[t]}
\put(243.50,81.50){\oval(3.00,3.00)[b]}
\put(246.50,81.50){\oval(3.00,3.00)[t]}
\put(249.50,81.50){\oval(3.00,3.00)[b]}
\put(252.50,81.50){\oval(3.00,3.00)[t]}
\put(239.10,101.00){\line(0,-1){20.00}}
\put(274.00,101.00){\line(-1,0){14.91}}
\put(274.00,81.00){\line(-1,0){14.91}}
\put(229.00,101.00){\vector(1,0){0.98}}
\put(230.00,81.00){\vector(-1,0){1.01}}
\put(265.00,101.00){\vector(1,0){1.05}}
\put(267.00,81.00){\vector(-1,0){0.94}}
\put(56.00,91.00){\makebox(0,0)[cc]{$(9)$}}
\put(130.00,91.00){\makebox(0,0)[cc]{$(10)$}}
\put(197.00,91.00){\makebox(0,0)[cc]{$(11)$}}
\put(273.00,91.00){\makebox(0,0)[cc]{$(12)$}}
\
\put(34.00,35.00){\oval(3.00,3.00)[l]}
\put(34.00,38.00){\oval(3.00,3.00)[r]}
\put(34.00,41.00){\oval(3.00,3.00)[l]}
\put(11.00,73.00){\line(4,-5){20.00}}
\put(37.00,48.00){\line(4,5){20.00}}
\put(31.14,26.00){\line(-1,-1){20.00}}
\put(37.14,26.00){\line(1,-1){20.00}}
\put(46.00,42.00){\makebox(0,0)[cc]{$(13)$}}
\put(34.00,46.00){\circle{6.00}}
\put(23.00,57.50){\vector(1,-1){0.97}}
\put(44.00,57.00){\vector(1,1){0.97}}
\put(34.00,51.00){\oval(3.00,3.00)[l]}
\put(34.00,54.00){\oval(3.00,3.00)[r]}
\put(34.00,57.00){\oval(3.00,3.00)[l]}
\put(34.00,60.00){\oval(3.00,3.00)[r]}
\put(34.00,63.00){\oval(3.00,3.00)[l]}
\put(34.00,66.00){\oval(3.00,3.00)[r]}
\put(31.00,26.00){\framebox(6.00,7.00)[cc]{}}
\put(24.00,19.00){\vector(-1,-1){0.97}}
\put(48.00,15.00){\vector(-1,1){1.02}}
\
\put(107.00,34.00){\oval(3.00,3.00)[l]}
\put(107.00,37.00){\oval(3.00,3.00)[r]}
\put(107.00,40.00){\oval(3.00,3.00)[l]}
\put(107.00,9.00){\oval(3.00,3.00)[l]}
\put(107.00,12.00){\oval(3.00,3.00)[r]}
\put(107.00,15.00){\oval(3.00,3.00)[l]}
\put(107.00,18.00){\oval(3.00,3.00)[r]}
\put(107.00,21.00){\oval(3.00,3.00)[l]}
\put(107.00,24.00){\oval(3.00,3.00)[r]}
\put(84.00,74.00){\line(4,-5){20.00}}
\put(110.00,49.00){\line(4,5){20.00}}
\put(104.14,27.00){\line(-1,-1){20.00}}
\put(109.14,27.00){\line(1,-1){20.00}}
\put(126.14,10.00){\vector(-1,1){9.00}}
\put(94.00,61.50){\vector(3,-4){1.67}}
\put(119.00,60.25){\vector(3,4){1.67}}
\put(119.00,43.00){\makebox(0,0)[cc]{$(14)$}}
\put(107.00,29.00){\circle{6.00}}
\put(104.00,42.00){\framebox(6.00,7.00)[cc]{}}
\put(96.00,19.00){\vector(-1,-1){0.95}}
\
\put(171.00,41.00){\oval(3.00,3.00)[b]}
\put(174.00,41.00){\oval(3.00,3.00)[t]}
\put(177.00,41.00){\oval(3.00,3.00)[b]}
\put(205.14,64.00){\line(-1,-1){20.00}}
\put(185.14,38.00){\line(1,-1){20.00}}
\put(204.81,46.00){\makebox(0,0)[cc]{$(15)$}}
\put(223.14,65.00){\line(1,-1){20.00}}
\put(243.14,39.00){\line(-1,-1){20.00}}
\put(283.81,47.00){\makebox(0,0)[cc]{$(16)$}}
\put(166.00,41.00){\circle{6.00}}
\put(164.00,44.00){\line(-1,1){16.93}}
\put(164.00,38.00){\line(-1,-1){16.93}}
\put(156.00,30.00){\vector(-1,-1){0.98}}
\put(156.00,52.00){\vector(1,-1){1.04}}
\put(194.00,53.00){\vector(1,1){1.04}}
\put(262.00,43.00){\circle{6.00}}
\put(263.00,40.00){\line(0,0){0.00}}
\put(264.00,45.00){\line(1,1){17.01}}
\put(264.00,41.00){\line(5,-6){15.82}}
\put(234.00,30.00){\vector(-1,-1){1.05}}
\put(271.00,32.00){\vector(-1,2){1.00}}
\put(272.00,53.00){\vector(1,1){0.97}}
\put(166.00,46.00){\oval(3.00,3.00)[l]}
\put(166.00,49.00){\oval(3.00,3.00)[r]}
\put(166.00,52.00){\oval(3.00,3.00)[l]}
\put(166.00,55.00){\oval(3.00,3.00)[r]}
\put(166.00,58.00){\oval(3.00,3.00)[l]}
\put(166.00,61.00){\oval(3.00,3.00)[r]}
\put(251.00,42.00){\oval(3.00,3.00)[b]}
\put(254.00,42.00){\oval(3.00,3.00)[t]}
\put(257.00,42.00){\oval(3.00,3.00)[b]}
\put(262.00,48.00){\oval(3.00,3.00)[l]}
\put(262.00,51.00){\oval(3.00,3.00)[r]}
\put(262.00,54.00){\oval(3.00,3.00)[l]}
\put(262.00,57.00){\oval(3.00,3.00)[r]}
\put(262.00,60.00){\oval(3.00,3.00)[l]}
\put(262.00,63.00){\oval(3.00,3.00)[r]}
\put(179.00,38.00){\framebox(6.00,6.00)[cc]{}}
\put(194.00,29.00){\vector(-1,1){1.06}}
\put(243.50,39.00){\framebox(6.00,6.00)[cc]{}}
\put(234.00,54.00){\vector(1,-1){0.93}}
\end{picture}

\noindent
{\small \bf Fig.~1: }{
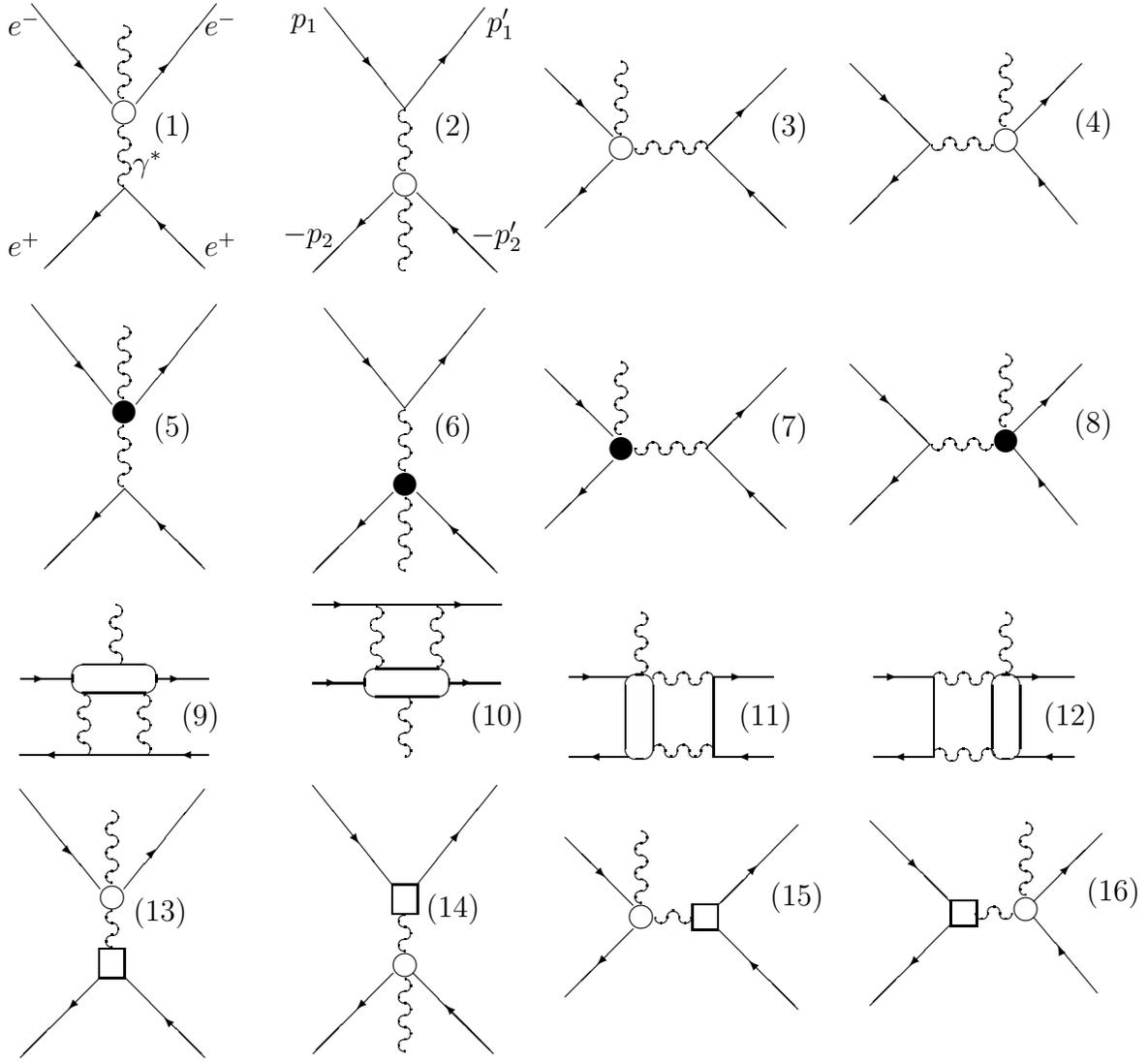
G and B type Feynman diagrams for radiative Bhabha scattering.}
\hspace{0.5cm}

\newpage

\hspace*{-1cm}
\unitlength=0.52mm
\special{em:linewidth 0.8pt}
\linethickness{0.8pt}
\begin{picture}(356.00,178.00)
\put(36.00,134.00){\oval(3.00,3.00)[l]}
\put(36.00,137.00){\oval(3.00,3.00)[r]}
\put(36.00,140.00){\oval(3.00,3.00)[l]}
\put(36.00,143.00){\oval(3.00,3.00)[r]}
\put(36.00,146.00){\oval(3.00,3.00)[l]}
\put(13.00,178.00){\line(4,-5){20.00}}
\put(39.00,153.00){\line(4,5){20.00}}
\put(125.00,124.00){\makebox(0,0)[cc]{$(a)$}}
\put(36.00,151.00){\circle{6.00}}
\put(25.00,162.50){\vector(1,-1){0.97}}
\put(46.00,162.00){\vector(1,1){0.97}}
\put(36.00,156.00){\oval(3.00,3.00)[l]}
\put(36.00,159.00){\oval(3.00,3.00)[r]}
\put(36.00,162.00){\oval(3.00,3.00)[l]}
\put(36.00,165.00){\oval(3.00,3.00)[r]}
\put(36.00,168.00){\oval(3.00,3.00)[l]}
\put(36.00,171.00){\oval(3.00,3.00)[r]}
\
\put(36.00,66.00){\oval(3.00,3.00)[l]}
\put(36.00,69.00){\oval(3.00,3.00)[r]}
\put(36.00,72.00){\oval(3.00,3.00)[l]}
\put(36.00,75.00){\oval(3.00,3.00)[r]}
\put(36.00,78.00){\oval(3.00,3.00)[l]}
\put(13.00,110.00){\line(4,-5){20.00}}
\put(39.00,85.00){\line(4,5){20.00}}
\put(153.00,56.00){\makebox(0,0)[cc]{$(d)$}}
\put(36.00,83.00){\circle*{6.00}}
\put(25.00,94.50){\vector(1,-1){0.97}}
\put(46.00,94.00){\vector(1,1){0.97}}
\put(36.00,88.00){\oval(3.00,3.00)[l]}
\put(36.00,91.00){\oval(3.00,3.00)[r]}
\put(36.00,94.00){\oval(3.00,3.00)[l]}
\put(36.00,97.00){\oval(3.00,3.00)[r]}
\put(36.00,100.00){\oval(3.00,3.00)[l]}
\put(36.00,103.00){\oval(3.00,3.00)[r]}
\put(67.00,149.00){\makebox(0,0)[cc]{$=$}}
\put(67.00,81.00){\makebox(0,0)[cc]{$=$}}
\put(93.00,136.00){\oval(3.00,3.00)[l]}
\put(93.00,139.00){\oval(3.00,3.00)[r]}
\put(93.00,142.00){\oval(3.00,3.00)[l]}
\put(93.00,145.00){\oval(3.00,3.00)[r]}
\put(93.00,148.00){\oval(3.00,3.00)[l]}
\put(73.00,175.00){\line(4,-5){20.00}}
\put(93.00,150.00){\line(4,5){20.00}}
\put(93.00,66.00){\oval(3.00,3.00)[l]}
\put(93.00,69.00){\oval(3.00,3.00)[r]}
\put(93.00,72.00){\oval(3.00,3.00)[l]}
\put(93.00,75.00){\oval(3.00,3.00)[r]}
\put(93.00,78.00){\oval(3.00,3.00)[l]}
\put(83.00,98.00){\oval(3.00,3.00)[r]}
\put(83.00,101.00){\oval(3.00,3.00)[l]}
\put(83.00,104.00){\oval(3.00,3.00)[r]}
\put(83.00,107.00){\oval(3.00,3.00)[l]}
\put(77.00,170.00){\vector(1,-1){1.03}}
\put(88.00,156.00){\vector(1,-1){0.97}}
\put(86.00,161.00){\oval(3.00,3.00)[r]}
\put(86.00,164.00){\oval(3.00,3.00)[l]}
\put(86.00,167.00){\oval(3.00,3.00)[r]}
\put(86.00,170.00){\oval(3.00,3.00)[l]}
\put(86.00,173.00){\oval(3.00,3.00)[r]}
\put(153.00,136.00){\oval(3.00,3.00)[l]}
\put(153.00,139.00){\oval(3.00,3.00)[r]}
\put(153.00,142.00){\oval(3.00,3.00)[l]}
\put(153.00,145.00){\oval(3.00,3.00)[r]}
\put(153.00,148.00){\oval(3.00,3.00)[l]}
\put(133.00,175.00){\line(4,-5){20.00}}
\put(153.00,150.00){\line(4,5){20.00}}
\put(162.00,164.00){\oval(3.00,3.00)[l]}
\put(162.00,167.00){\oval(3.00,3.00)[r]}
\put(162.00,170.00){\oval(3.00,3.00)[l]}
\put(162.00,173.00){\oval(3.00,3.00)[r]}
\put(125.00,149.00){\makebox(0,0)[cc]{$+$}}
\put(168.00,169.00){\vector(1,1){0.99}}
\put(156.00,154.00){\vector(1,1){1.04}}
\put(144.00,161.00){\vector(1,-1){0.99}}
\put(93.00,81.00){\oval(3.00,3.00)[r]}
\put(125.00,81.00){\makebox(0,0)[cc]{$+$}}
\put(155.00,68.00){\oval(3.00,3.00)[l]}
\put(155.00,71.00){\oval(3.00,3.00)[r]}
\put(155.00,74.00){\oval(3.00,3.00)[l]}
\put(155.00,77.00){\oval(3.00,3.00)[r]}
\put(155.00,80.00){\oval(3.00,3.00)[l]}
\put(141.00,100.00){\oval(3.00,3.00)[r]}
\put(141.00,103.00){\oval(3.00,3.00)[l]}
\put(141.00,106.00){\oval(3.00,3.00)[r]}
\put(155.00,83.00){\oval(3.00,3.00)[r]}
\put(182.00,81.00){\makebox(0,0)[cc]{$+$}}
\put(211.00,66.00){\oval(3.00,3.00)[l]}
\put(211.00,69.00){\oval(3.00,3.00)[r]}
\put(211.00,72.00){\oval(3.00,3.00)[l]}
\put(211.00,75.00){\oval(3.00,3.00)[r]}
\put(211.00,78.00){\oval(3.00,3.00)[l]}
\put(191.00,108.00){\line(4,-5){20.00}}
\put(201.00,98.00){\oval(3.00,3.00)[r]}
\put(201.00,101.00){\oval(3.00,3.00)[l]}
\put(201.00,104.00){\oval(3.00,3.00)[r]}
\put(201.00,107.00){\oval(3.00,3.00)[l]}
\put(194.00,104.00){\vector(1,-1){1.00}}
\put(211.00,81.00){\oval(3.00,3.00)[r]}
\put(237.00,81.00){\makebox(0,0)[cc]{$+$}}
\put(265.00,66.00){\oval(3.00,3.00)[l]}
\put(265.00,69.00){\oval(3.00,3.00)[r]}
\put(265.00,72.00){\oval(3.00,3.00)[l]}
\put(265.00,75.00){\oval(3.00,3.00)[r]}
\put(265.00,78.00){\oval(3.00,3.00)[l]}
\put(245.00,108.00){\line(4,-5){20.00}}
\put(265.00,83.00){\line(4,5){20.00}}
\put(272.00,92.00){\vector(1,1){0.97}}
\put(260.00,91.50){\oval(3.00,3.00)[r]}
\put(260.00,94.50){\oval(3.00,3.00)[l]}
\put(260.00,97.50){\oval(3.00,3.00)[r]}
\put(260.00,100.50){\oval(3.00,3.00)[l]}
\put(265.00,81.00){\oval(3.00,3.00)[r]}
\put(291.00,81.00){\makebox(0,0)[cc]{$+$}}
\put(204.50,93.50){\oval(3.00,3.00)[t]}
\put(207.50,93.50){\oval(3.00,3.00)[b]}
\put(210.50,93.50){\oval(3.00,3.00)[t]}
\put(213.50,93.50){\oval(3.00,3.00)[b]}
\put(216.50,93.50){\oval(3.00,3.00)[t]}
\put(256.50,96.50){\oval(3.00,3.00)[t]}
\put(259.50,96.50){\oval(3.00,3.00)[b]}
\put(262.50,96.50){\oval(3.00,3.00)[t]}
\put(265.50,96.50){\oval(3.00,3.00)[b]}
\put(268.50,96.50){\oval(3.00,3.00)[t]}
\put(271.50,96.50){\oval(3.00,3.00)[b]}
\put(274.50,96.50){\oval(3.00,3.00)[t]}
\put(260.00,104.00){\oval(3.00,3.00)[r]}
\put(260.00,107.00){\oval(3.00,3.00)[l]}
\put(248.00,104.00){\vector(1,-1){1.06}}
\put(331.00,81.00){\makebox(0,0)[cc]{4 mirror FD}}
\put(36.50,34.50){\oval(29.00,7.00)[]}
\put(22.00,34.00){\line(-1,0){13.92}}
\put(65.00,34.00){\line(-1,0){13.92}}
\put(36.50,40.00){\oval(3.00,3.00)[l]}
\put(36.50,43.00){\oval(3.00,3.00)[r]}
\put(36.50,46.00){\oval(3.00,3.00)[l]}
\put(36.50,49.00){\oval(3.00,3.00)[r]}
\put(25.00,17.50){\oval(3.00,3.00)[l]}
\put(25.00,20.50){\oval(3.00,3.00)[r]}
\put(25.00,23.50){\oval(3.00,3.00)[l]}
\put(25.00,26.50){\oval(3.00,3.00)[r]}
\put(25.00,29.50){\oval(3.00,3.00)[l]}
\put(49.00,17.50){\oval(3.00,3.00)[l]}
\put(49.00,20.50){\oval(3.00,3.00)[r]}
\put(49.00,23.50){\oval(3.00,3.00)[l]}
\put(49.00,26.50){\oval(3.00,3.00)[r]}
\put(49.00,29.50){\oval(3.00,3.00)[l]}
\put(71.00,29.00){\makebox(0,0)[cc]{$=$}}
\put(166.00,6.00){\makebox(0,0)[cc]{$(b)$}}
\put(92.00,20.50){\oval(3.00,3.00)[l]}
\put(92.00,23.50){\oval(3.00,3.00)[r]}
\put(92.00,26.50){\oval(3.00,3.00)[l]}
\put(92.00,29.50){\oval(3.00,3.00)[r]}
\put(92.00,32.50){\oval(3.00,3.00)[l]}
\put(111.00,20.50){\oval(3.00,3.00)[l]}
\put(111.00,23.50){\oval(3.00,3.00)[r]}
\put(111.00,26.50){\oval(3.00,3.00)[l]}
\put(111.00,29.50){\oval(3.00,3.00)[r]}
\put(111.00,32.50){\oval(3.00,3.00)[l]}
\put(86.50,36.00){\oval(3.00,3.00)[l]}
\put(86.50,39.00){\oval(3.00,3.00)[r]}
\put(86.50,42.00){\oval(3.00,3.00)[l]}
\put(86.50,45.00){\oval(3.00,3.00)[r]}
\put(86.50,48.00){\oval(3.00,3.00)[l]}
\put(13.00,34.00){\vector(1,0){1.02}}
\put(57.00,34.00){\vector(1,0){0.96}}
\put(82.00,34.00){\vector(1,0){1.01}}
\put(99.00,34.00){\vector(1,0){1.04}}
\put(115.00,34.00){\vector(1,0){0.95}}
\put(206.00,89.00){\vector(1,-1){0.94}}
\put(280.00,102.00){\vector(1,1){0.94}}
\put(78.00,34.00){\line(1,0){46.00}}
\put(133.00,29.00){\makebox(0,0)[cc]{$+$}}
\put(157.00,20.50){\oval(3.00,3.00)[l]}
\put(157.00,23.50){\oval(3.00,3.00)[r]}
\put(157.00,26.50){\oval(3.00,3.00)[l]}
\put(157.00,29.50){\oval(3.00,3.00)[r]}
\put(157.00,32.50){\oval(3.00,3.00)[l]}
\put(176.00,20.50){\oval(3.00,3.00)[l]}
\put(176.00,23.50){\oval(3.00,3.00)[r]}
\put(176.00,26.50){\oval(3.00,3.00)[l]}
\put(176.00,29.50){\oval(3.00,3.00)[r]}
\put(176.00,32.50){\oval(3.00,3.00)[l]}
\put(165.50,36.00){\oval(3.00,3.00)[l]}
\put(165.50,39.00){\oval(3.00,3.00)[r]}
\put(165.50,42.00){\oval(3.00,3.00)[l]}
\put(165.50,45.00){\oval(3.00,3.00)[r]}
\put(165.50,48.00){\oval(3.00,3.00)[l]}
\put(147.00,34.00){\vector(1,0){1.01}}
\put(180.00,34.00){\vector(1,0){0.95}}
\put(143.00,34.00){\line(1,0){46.00}}
\put(228.00,20.50){\oval(3.00,3.00)[l]}
\put(228.00,23.50){\oval(3.00,3.00)[r]}
\put(228.00,26.50){\oval(3.00,3.00)[l]}
\put(228.00,29.50){\oval(3.00,3.00)[r]}
\put(228.00,32.50){\oval(3.00,3.00)[l]}
\put(247.00,20.50){\oval(3.00,3.00)[l]}
\put(247.00,23.50){\oval(3.00,3.00)[r]}
\put(247.00,26.50){\oval(3.00,3.00)[l]}
\put(247.00,29.50){\oval(3.00,3.00)[r]}
\put(247.00,32.50){\oval(3.00,3.00)[l]}
\put(251.50,36.00){\oval(3.00,3.00)[l]}
\put(251.50,39.00){\oval(3.00,3.00)[r]}
\put(251.50,42.00){\oval(3.00,3.00)[l]}
\put(251.50,45.00){\oval(3.00,3.00)[r]}
\put(251.50,48.00){\oval(3.00,3.00)[l]}
\put(218.00,34.00){\vector(1,0){1.01}}
\put(236.00,34.00){\vector(1,0){1.04}}
\put(214.00,34.00){\line(1,0){46.00}}
\put(201.00,29.00){\makebox(0,0)[cc]{$+$}}
\put(255.00,34.00){\vector(1,0){0.98}}
\put(270.00,29.00){\makebox(0,0)[cc]{$+$}}
\put(295.00,29.00){\makebox(0,0)[cc]{$(1\leftrightarrow 2)$}}
\put(104.00,97.00){\vector(1,1){0.95}}
\put(105.00,165.00){\vector(1,1){1.05}}
\put(215.00,134.00){\oval(3.00,3.00)[l]}
\put(215.00,137.00){\oval(3.00,3.00)[r]}
\put(215.00,140.00){\oval(3.00,3.00)[l]}
\put(215.00,143.00){\oval(3.00,3.00)[r]}
\put(215.00,146.00){\oval(3.00,3.00)[l]}
\put(192.00,178.00){\line(4,-5){20.00}}
\put(218.00,153.00){\line(4,5){20.00}}
\put(276.00,124.00){\makebox(0,0)[cc]{$(f)$}}
\put(225.00,162.00){\vector(1,1){0.97}}
\put(246.00,149.00){\makebox(0,0)[cc]{$=$}}
\put(212.00,148.00){\framebox(6.00,6.00)[cc]{}}
\put(203.00,164.00){\vector(1,-1){1.06}}
\put(92.00,14.00){\makebox(0,0)[cc]{$1$}}
\put(111.00,14.00){\makebox(0,0)[cc]{$2$}}
\put(157.00,14.00){\makebox(0,0)[cc]{$1$}}
\put(176.00,14.00){\makebox(0,0)[cc]{$2$}}
\put(228.00,14.00){\makebox(0,0)[cc]{$1$}}
\put(247.00,14.00){\makebox(0,0)[cc]{$2$}}
\put(295.00,6.00){\makebox(0,0)[cc]{$(c)$}}
\put(331.00,57.00){\makebox(0,0)[cc]{$(e)$}}
\put(276.00,145.00){\oval(3.00,3.00)[r]}
\put(276.00,148.00){\oval(3.00,3.00)[l]}
\put(256.00,175.00){\line(4,-5){20.00}}
\put(276.00,150.00){\line(4,5){20.00}}
\put(336.00,136.00){\oval(3.00,3.00)[l]}
\put(336.00,139.00){\oval(3.00,3.00)[r]}
\put(336.00,142.00){\oval(3.00,3.00)[l]}
\put(336.00,145.00){\oval(3.00,3.00)[r]}
\put(336.00,148.00){\oval(3.00,3.00)[l]}
\put(316.00,175.00){\line(4,-5){20.00}}
\put(336.00,150.00){\line(4,5){20.00}}
\put(308.00,149.00){\makebox(0,0)[cc]{$+$}}
\put(351.00,169.00){\vector(1,1){0.99}}
\put(327.00,161.00){\vector(1,-1){0.99}}
\put(288.00,165.00){\vector(1,1){1.05}}
\put(276.00,140.50){\circle{6.00}}
\put(276.00,135.50){\oval(3.00,3.00)[l]}
\put(276.00,132.50){\oval(3.00,3.00)[r]}
\put(327.50,163.50){\oval(3.00,3.00)[t]}
\put(330.50,163.50){\oval(3.00,3.00)[b]}
\put(333.50,163.50){\oval(3.00,3.00)[t]}
\put(336.50,163.50){\oval(3.00,3.00)[b]}
\put(339.50,163.50){\oval(3.00,3.00)[t]}
\put(342.50,163.50){\oval(3.00,3.00)[b]}
\put(345.50,163.50){\oval(3.00,3.00)[t]}
\put(320.00,170.00){\vector(1,-1){1.01}}
\put(337.00,151.00){\vector(2,3){2.00}}
\put(337.00,124.00){\makebox(0,0)[cc]{$(g)$}}
\put(336.00,133.00){\oval(3.00,3.00)[r]}
\put(267.00,161.00){\vector(1,-1){1.01}}
\put(183.00,150.00){\makebox(0,0)[cc]{$;$}}
\
\put(78.00,99.20){\oval(3.00,3.00)[l]}
\put(78.00,96.20){\oval(3.00,3.00)[r]}
\put(78.00,93.20){\oval(3.00,3.00)[l]}
\put(80.00,91.20){\oval(3.00,3.00)[b]}
\put(83.00,91.20){\oval(3.00,3.00)[t]}
\put(86.00,91.20){\oval(3.00,3.00)[b]}
\
\put(93.00,86.00){\line(-1,1){21.00}}
\put(93.00,86.00){\line(1,1){20.97}}
\put(93.00,84.00){\oval(3.00,3.00)[l]}
\put(74.00,105.00){\vector(1,-1){1.01}}
\put(143.00,94.20){\oval(3.00,3.00)[l]}
\put(143.00,91.20){\oval(3.00,3.00)[r]}
\put(143.00,88.20){\oval(3.00,3.00)[l]}
\put(145.00,86.20){\oval(3.00,3.00)[b]}
\put(148.00,86.20){\oval(3.00,3.00)[t]}
\put(151.00,86.20){\oval(3.00,3.00)[b]}
\put(155.00,84.00){\line(-1,1){24.01}}
\put(155.00,84.00){\line(5,6){19.99}}
\put(135.00,104.00){\vector(1,-1){1.02}}
\put(149.00,90.00){\vector(1,-1){0.95}}
\put(165.00,96.00){\vector(1,1){0.98}}
\put(89.00,90.00){\vector(1,-1){2.06}}
\put(159.00,34.00){\vector(1,0){1.94}}
\put(168.00,34.00){\vector(1,0){2.01}}
\put(211.00,83.00){\line(2,3){16.74}}
\put(221.00,98.00){\vector(2,3){2.05}}
\put(213.00,86.00){\vector(2,3){1.97}}
\put(260.00,89.00){\vector(3,-4){3.00}}
\end{picture}

\noindent
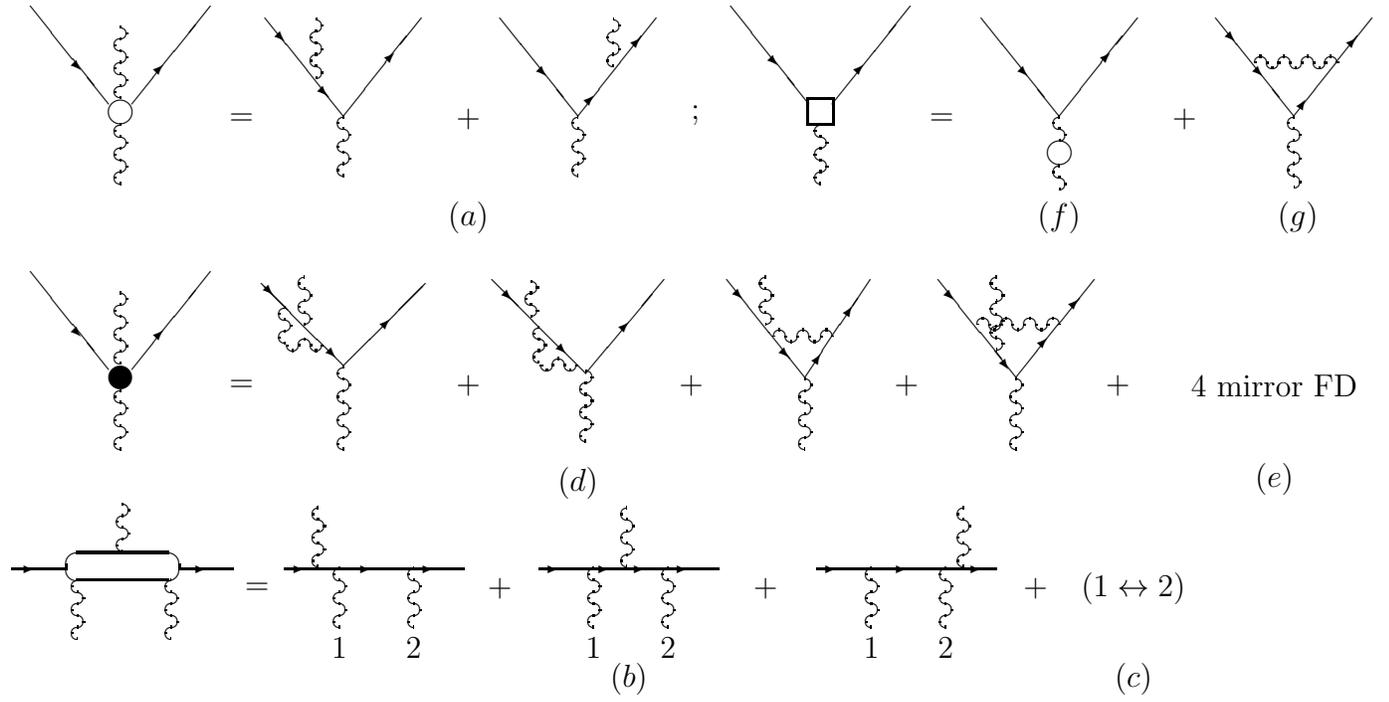
{\small\bf Fig.~2: }{ Content of the notation for Fig.~1. }

\end{document}